\documentclass[twocolumn, 11pt]{aastex631}
\usepackage{amsmath}
\usepackage{layouts}

\begin{document}

\title{Self-Similar Mass Accretion History in Scale-Free Simulations}

\author{John Soltis}
\affiliation{Department of Physics \& Astronomy, Johns Hopkins University, Baltimore, MD 21218, USA}
\affiliation{Center for Computational Astrophysics, Flatiron Institute, 162 Fifth Ave., New York, NY 10010}
\affiliation{Scientific Computing Core, Flatiron Institute, 162 Fifth Ave., New York, NY 10010}

\author{Lehman H.~Garrison}
\affiliation{Scientific Computing Core, Flatiron Institute, 162 Fifth Ave., New York, NY 10010}

\begin{abstract}
    Using a scale-free $N$-body simulation generated with the \textsc{Abacus} $N$-body code, we test the robustness of halo mass accretion histories via their convergence to self-similarity. We compare two halo finders, \textsc{Rockstar} and \textsc{CompaSO}. We find superior self-similarity in halo mass accretion histories determined using \textsc{Rockstar}, with convergence to 5\% or better between $\sim10^2$ to $10^5$ particles. For \textsc{CompaSO} we find weaker convergence over a similar region, with at least 10\% between $\sim10^2$ to $10^4$ particles. Furthermore, we find the convergence to self-similarity improves as the simulation evolves, with the largest and deepest regions of convergence appearing after the scale factor quadrupled from the time at which non-linear structures begin to form. With sufficient time evolution, halo mass accretion histories are converged to self-similarity within 5\% with as few as $\sim70$ particles for \textsc{CompaSO} and within 2\% for as few as $\sim30$ particles for \textsc{Rockstar}.
\end{abstract}

\section{Introduction}\label{introduction}
In our own Universe, the mass accretion history of halos plays an important role in the dynamical state of galaxy clusters \citep[for reviews, see][]{Molnar_2016, Pratt_2019}. These massive virialized objects can provide a useful probe of cosmology, however their dynamical state can bias cosmological parameter estimation \citep[e.g.,][]{Lau_2009, Nelson_2014, Shi_2015, Gianfagna_2021}. Similarly, galaxy clusters are a useful laboratory for studying dark matter \citep[e.g.,][]{Eckert_2022}. Here dynamical state, and thus mass accretion history, sometimes provides a more salutary role than with cosmological parameter constraints. In some cases, most famously the Bullet Cluster \citep{Clowe_2006}, an extreme dynamical state allows us to separately observe the dark matter and gas components. Mass accretion history is also thought to play an important role in the evolution of galaxies in clusters, with infalling galaxies losing gas to the surrounding intracluster medium \citep[for a review, see][]{Boselli_2022}.

Given the importance of halo mass accretion history in understanding cosmology, dark matter, and astrophysics, it is no surprise that there has been a great deal of interest in better understanding mass accretion history itself. Links have been found between mass accretion and a number of other important cluster properties, like concentration, asymmetry, the position of the brightest core galaxy relative to the X-ray centroid, and the splashback radius \citep[e.g.,][]{Lovisari_2017, Dupourque_2022, Shin_2023}. Efforts have been made to construct mass accretion or dynamical state proxies out of the observable morphology of galaxy clusters, or to estimate it more directly (Soltis et al in prep.; \citealt{De_Luca_2021, Capalbo_2022, Perez_2023, Pizzardo_2023, Arendt_2024}). Research into the relationship between mass accretion history and cosmological parameters is also ongoing (Warburton et al. in prep.; \citealt{Amoura_2024}).

To draw a link between mass accretion histories and observable properties of clusters, or to choices of cosmology, one typically needs to use cosmological simulations (see \citealt{Pizzardo_2022} for a near exception, although $N$-body simulations were still used to inform and validate the method). Applying conclusions drawn from simulations faces an important challenge, however. What if the simulation is not physically accurate? How can we guarantee that the distribution of mass accretion histories in a simulation is representative of our Universe if we are unable to directly observe mass accretion history in our Universe? In this paper, we take an important step towards answering this question. 

Previous work has also sought to answer this question. For example, \citet{Mansfield_2021} examined the convergence of mass accretion rates in halos between lower and higher resolution dark matter only simulations. While this form of test is essential, we opt for a different, but complimentary, approach. We probe whether a simulation and halo finder deviate from expected behavior without relying on higher resolution simulations. This is important, because comparing to a higher resolution simulation can, at most, provide a relative test of convergence. However, higher resolution simulations are not guaranteed to be more converged to physically realistic results than lower resolution simulations.

Scale-free cosmologies present an ideal testing ground for evaluating mass accretion histories. In a scale-free cosmology, we expect any dimensionless physical property to behave self-similarly (see Section \ref{scale_free} for more details). Where a property deviates from self-similarity, we can say that it is dependent on non-physical properties of the simulation, or halo-finder, and is therefore not physically accurate. This presents a useful, and more absolute, probe of the robustness of halo mass accretion histories than can be attained by comparing simulations of different resolutions. Applying the results of this testing to other simulations and cosmologies is another challenge. As will be discussed in sections \ref{Abacus} and \ref{LCDM}, one can use the choice of spectral index to apply the results to other cosmologies. Moreover, future exploration of different $N$-body codes and spectral indices could provide upper bounds on the range of reliability of mass accretion histories, or other properties, for different simulations and use cases. 

It is important to emphasize that this scale-free test probes the accuracy of simulations in only a negative sense. Deviations from self-similarity suggest that a simulation and/or halo finder is unreliable. The presence of self-similarity, however, does not guarantee that the examined properties are physically realistic. Constraints from this test thus serve as an upper bound on the robustness of simulations.

Several papers have been written on using the \textsc{Abacus} $N$-body code and scale-free simulations. In \citet{Joyce_2021}, the self-similarity of the two point correlation function of particles in a scale-free simulation generated using \textsc{Abacus} is tested. In using this test to quantify the resolution of the simulation, the authors find that the length scales on which they observed convergence to self-similarity propagated from larger to smaller scales. This is a result that we will also observe. \citet{Leroy_2021} applies the same technique to investigate the convergence to self-similarity of the halo mass function and the halo-halo correlation function. In doing so they compare the performance of two different halo finders, a friends-of-friends algorithm and \textsc{Rockstar} \citep{Rockstar}. They observe good convergence using \textsc{Rockstar} and very limited convergence using the friends-of-friends algorithm. \citet{Garrison_2021_softening} once again uses the two point correlation function of particles in a scale-free simulation generated using \textsc{Abacus}, this time to test the impact of the softening scheme used. In addition to observing that convergence to self-similarity improves at smaller scales at late times, the analysis also confirmed the fiducial choice of softening regime (recapped briefly in Section \ref{Abacus}). \citet{Garrison_2022_k_nearest} use a $k$-nearest neighbor probability distribution to assess convergence. In doing so they probe the density at different particle and density scales, finding that spheres containing only 32 particles will be converged at densities typical of halos. 

Our work most closely builds off the methods and analysis of a series of papers, \citet{Maleubre_2022}, \citet{Maleubre_2023}, and \citet{Maleubre_2024}. \citet{Maleubre_2022} analyzed the convergence of the matter power spectrum, while \citet{Maleubre_2023} examined the convergence of radial pair-wise velocity of particles. Finally, \citet{Maleubre_2024} extends on the previous works by investigating the convergence of the halo mass function, two point correlation function, and the mean radial pair-wise velocity of halos. Similar to \citet{Leroy_2021} and exactly as in this work, \citet{Maleubre_2024} compares the performance of two halo finders, \textsc{Rockstar} \citep{Rockstar} and \textsc{CompaSO} \citep{CompaSO}. The results of that work are in line with previous results, namely that \textsc{Rockstar} performs better than other halo finders, and that convergence improves as the simulation evolves. 

In this paper we use the self-similarity property of scale-free cosmologies to probe the robustness of halo mass accretion histories. We use a scale-free simulation from the same family of \textsc{Abacus} simulations \citep{Garrison_2021_Abacus} used in previous work \citep[e.g,][]{Maleubre_2024}. Like \citet{Maleubre_2024}, we compare the performance of two halo finders, \textsc{Rockstar} \citep{Rockstar} and \textsc{CompaSO} \citep{CompaSO}. Our work serves as a natural extension of \citet{Maleubre_2024}, in so far as we use the same simulation, test the same halo finders, and that our analysis probes the convergence of the halo merger trees. In this work, we present bounds on self-similarity for both halo finders and the prospect for applying them to other simulations. 

This paper is structured as follows. In Section \ref{methods} we provide a brief overview of scale-free cosmologies (\ref{scale_free}), a description of our definition of mass accretion history (\ref{mah_def}), and explain our criteria for evaluating convergence to self-similarity (\ref{self_similarity}). In Section \ref{sims_and_finders} we discuss our choice of $N$-body code and simulation (\ref{Abacus}), and the two halo finders used (\ref{finders_descriptions}). Next, in Section \ref{results}, we describe our results for both halo finders (\ref{rockstar_results}, \ref{compaso_results}). Finally, we discuss the results in Section \ref{discussion} and conclude in Section \ref{conclusion}.

\section{Methodology}\label{methods}
\subsection{Scale-Free Cosmology}\label{scale_free}
$N$-body simulations with a scale-free cosmology have been used to study the behavior and robustness of halo properties, and the applicability of the stable clustering analytical approximation \citep[e.g.,][]{Efstathiou_1988, Colombi_1996, Cole_1996, Joyce_2007, Knollmann_2008, Elahi_2009}. Following the format of previous work \citep[][]{Joyce_2021, Leroy_2021, Garrison_2021_softening, Garrison_2022_k_nearest, Maleubre_2022, Maleubre_2023, Maleubre_2024}, we use the property of self-similarity in scale-free simulations to test the robustness of halo mass accretion histories. Below we briefly summarize scale-free simulations and self-similarity following the discussion in \citet{Maleubre_2024}. For further discussion of scale-free simulations and their properties, see \citet{Joyce_2021} and citations therein.

Scale-free cosmologies are Einstein-de Sitter universes, $\Omega_{\mathrm{Tot}} = \Omega_{M} = 1$, with a power law spectrum of initial fluctuations, $P(k) \propto k^{n}$, and a $a(t) \propto t^{2/3}$ expansion law. With these conditions, there exists only one relevant length scale, $R_{\mathrm{NL}}$, the scale at which the variance of the normalized linear amplitude of fluctuations, $\sigma_{\mathrm{lin}}$, is equal to one. This length scale is known as the scale of non-linearity, defined by
\begin{equation}\label{eqn:def_nonlinear_scale}
    \sigma_{\mathrm{lin}}(R_{\mathrm{NL}},a) = 1.
\end{equation}
From linear perturbation theory, we deduce that
\begin{equation}\label{eqn:nonlinear_evolution}
    R_{\mathrm{NL}} \propto a^{\frac{2}{3+n}}
\end{equation}
where $n$ is the spectral index. This suggests that, without the introduction of other length scales, the clustering must evolve self-similarly. Thus any dimensionless clustering statistic can be rescaled such that
\begin{equation}\label{eqn:rescaling}
    F(X,a) = F_0\left(\frac{X}{X_{\mathrm{NL}}(a)}\right).
\end{equation}
Here $X_{\mathrm{NL}}$ contains the time dependence of any quantity with the dimensions of $X$.

We can define $R_{\mathrm{NL}}$ in terms of the initial conditions of the simulation using equations \ref{eqn:def_nonlinear_scale} and \ref{eqn:nonlinear_evolution}. That is

\begin{equation}\label{eqn:R_def}
    R_{\mathrm{NL}}(a) = \ell \left(\frac{a}{a_i} \sigma_i \right)^{\frac{2}{3+n}}.
\end{equation}
Here $\sigma_i$ is variance of the normalized linear amplitude of fluctuations at $a_i$, where the initial particle spacing is set to $\ell$ (i.e., $\sigma_i \equiv \sigma_{\mathrm{lin}}(\ell, a_i)$). Using Equation \ref{eqn:R_def}, we can define a non-linear mass scale as
\begin{equation}\label{eqn:nonlinearmass}
    M_{\mathrm{NL}} = \frac{4 \pi}{3} \bar{\rho} R_{\mathrm{NL}}^3(a) = \frac{4 \pi}{3} M_{particle} \left(\sigma_i \frac{a}{a_i}\right)^{\frac{6}{3+n}}.
\end{equation}
We will use this non-linear mass scale to rescale our mass accretion history definition in the following subsections.

\subsection{Defining Halo Mass Accretion}\label{mah_def}

To take advantage of the self-similarity of dimensionless statistics in scale-free simulations, we must construct a dimensionless halo mass accretion history. Our definition is also constrained by the discrete temporal outputs of our simulation, so we define our mass accretion history in terms of the difference in the mass of halos and their descendants over two consecutive snapshots. Thus our mass accretion history can also be thought of as a mass accretion rate. With these constraints in mind, we define the halo mass accretion history of the $j$th cluster, $\Gamma_j$, as
\begin{equation} \label{eqn:mahofmofj}
\Gamma_j(a) = \frac{d \log(M)}{d \log(a)}
\end{equation}
As discussed, we must approximate the derivative in Equation \ref{eqn:mahofmofj} using the discrete snapshot outputs of the simulation. Conveniently, these snapshots are evenly spaced in $\log(a)$ (see Section \ref{Abacus} and Equation \ref{eqn:log_a_spacing}), thus we find that for the $j$th cluster, the mean mass accretion history at time $i$ is proportional to
\begin{equation} \label{eqn:approx_mahofmofj}
   \Gamma_j(a_i) \propto \frac{M_j(a_{i+1}) - M_j(a_{i})}{M_j(a_{i})}
\end{equation}
Therefore, the mass accretion history of the $j$th cluster is the difference between the mass of the $j$th halo at snapshot $i$, $M_j\left(a_{i}\right)$, and the mass of its most massive descendant at the following snapshot, $M_j\left(a_{i+1}\right)$, all normalized by $M_j\left(a_{i}\right)$.

We are not interested in the mass accretion rate of individual clusters, however, but in the population level behavior. Therefore, we define a mean mass accretion history, $\Gamma$, over the population of clusters at time $i$ with mass $M$ as
\begin{equation} \label{eqn:mahofm}
   \Gamma(M, a_i) = \left<\Gamma_j(a_i)\right> \propto \left<\frac{M_j(a_{i+1}) - M_j(a_{i})}{M_j(a_{i})}\right>
\end{equation}
where the angle brackets represent the average over halos whose masses satisfy
\begin{equation} \label{eqn:av_over_mahofm}
 M_j \in \left(M, M + d M\right).
\end{equation}
Note that this mean mass accretion history is now a function of both halo mass and scale factor.

Following the principles discussed in Section \ref{scale_free}, we rescale Equation \ref{eqn:mahofm} according to Equation \ref{eqn:rescaling}. In this case, the rescaling unit is the mass scale of non-linearity, seen in Equation \ref{eqn:nonlinearmass}. The resulting mean halo mass accretion history definition is
\begin{equation} \label{eqn:mahofMNL}
\begin{split}
\Gamma\left(\frac{M}{M_{\mathrm{NL}}}\right) & \propto \left<\frac{M_j(a_{i+1}) - M_j(a_{i})}{M_j(a_{i})}\right> \\
 \frac{M_j}{M_{\mathrm{NL}(a_i)}} & \in \left(\frac{M}{M_{\mathrm{NL}}}, \frac{M}{M_{\mathrm{NL}}} + d \frac{M}{M_{\mathrm{NL}}}\right).
\end{split}
\end{equation}
The only change between Equation \ref{eqn:mahofm} and \ref{eqn:mahofMNL} is that the halos in the self-similar mass accretion history definition are now averaged by their rescaled mass, $M/M_{\mathrm{NL}}$. The result of this is that $\Gamma$ is now only a function of this rescaled mass bin. If our simulation is perfectly scale-free, and our halo finder is not introducing any non-physical scales, we should not expect to see any additional time dependence. In practice, we observe deviations from self-similarity, and therefore our observed mean mass accretion history does have an additional time dependence (i.e., $\Gamma_{\mathrm{obs}}(M/M_{NL}, a)$).

\begin{figure*}
    \centering
    \fig{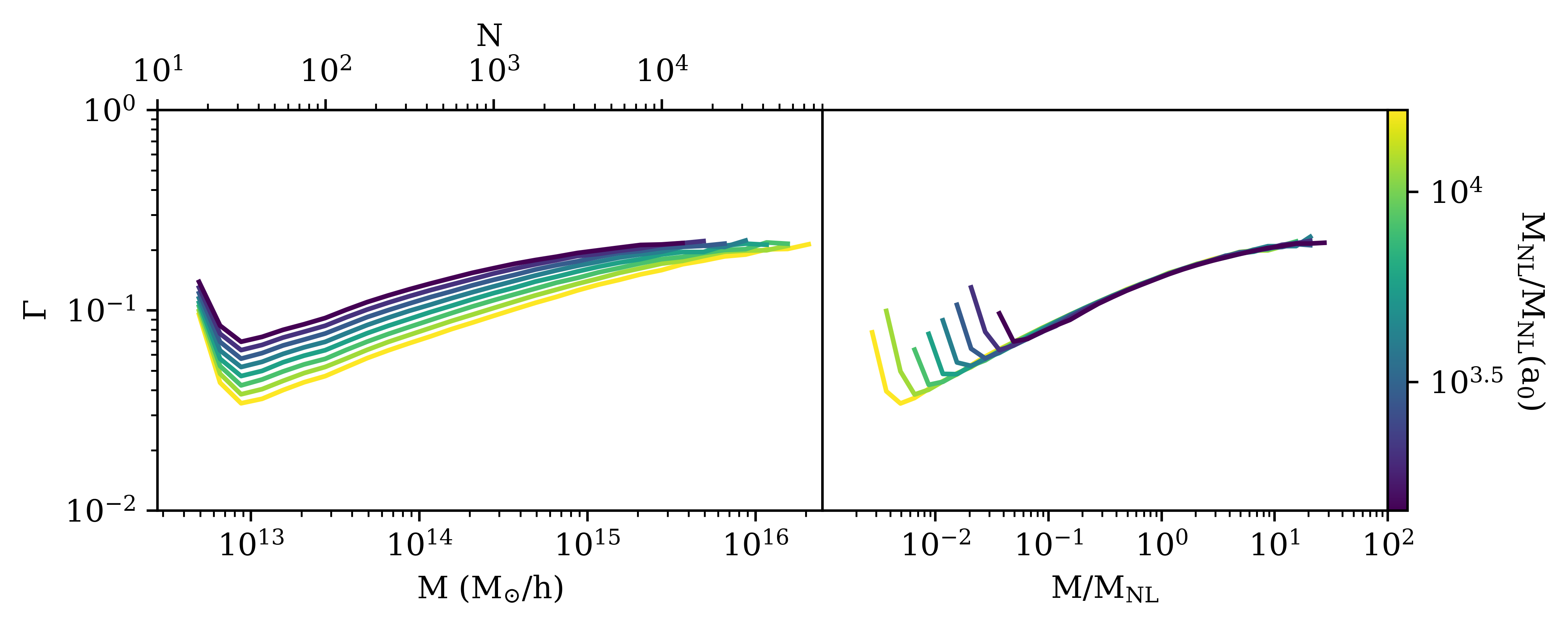}{\textwidth}{}
    \caption{\raggedright The mass accretion history, $\Gamma$, for the last 8 snapshots of our simulation of a scale-free model with $n=-1.5$, found using the \textsc{Rockstar} halo finder. The plot on the left shows the mean mass accretion history ($\Gamma$) versus the halo mass ($M$) and the halo particle count ($N$), where the mass and particle count are defined by the center of the mass bin used. The plot on the right shows the self-similar rescaling of $\Gamma$, now plotted versus the rescaled mass, $M/M_{\mathrm{NL}}$. Note that the lines in the $\Gamma$ vs $M/M_{\mathrm{NL}}$ plot now fall on top of each other near $M/M_{\mathrm{NL}} \sim 1$, indicating that $\Gamma$ is self-similar in that region. The color illustrates the time step of the lines in terms of the multiplicative change in the non-linear mass since $a_0$. For more information on the snapshot time steps, see Section \ref{Abacus} and Equation \ref{eqn:log_a_spacing}.}
    \label{fig:rescaling_example}
\end{figure*}

\begin{figure*}
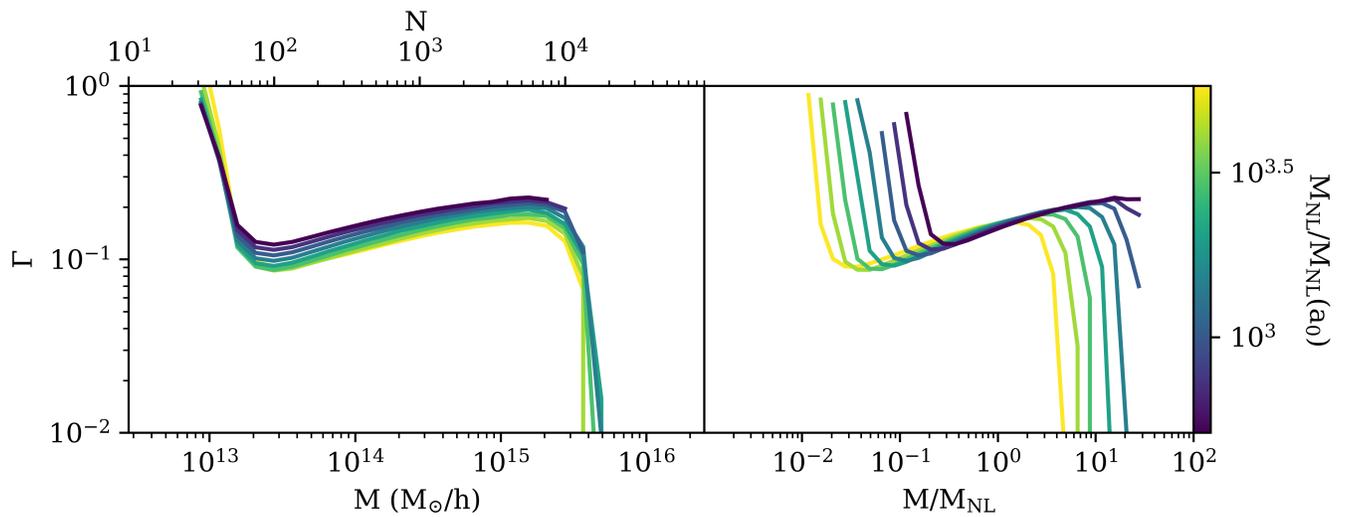

    \centering
    \fig{Paper_Plots_gtr20/CompaSO_n15_z_mean_past_joyce_mah_M200c_gtr20_vs_M_and_M_M_NL.png}{\textwidth}{}
    \caption{\raggedright Above is a similar figure to Figure \ref{fig:rescaling_example}, but for mass accretion histories calculated using \textsc{CompaSO}. Note, that as pointed out in Section \ref{compaso_results}, the last two snapshots were not available for the \textsc{CompaSO} data, so the last snapshot available is $\log_2(a/a_0) = 3.25$ instead of $3.5$. The behavior illustrated in this plot is similar to Figure \ref{fig:rescaling_example}, with self-similarity apparent at $M/M_{\mathrm{NL}}\sim 1$. However, unlike Figure \ref{fig:rescaling_example}, the the results diverge from self-similarity at late times. The difficulty the \textsc{CompaSO} halo finder has with high mass halos is discussed in sections \ref{compaso_results} and \ref{comparing_finders}.} 
    \label{fig:CompaSO_rescaling_example}
\end{figure*}

For our rescaled mass bins, we use 40 logarithmically spaced bins ranging $10^{-3}$ to $10^{2}$. This provides good coverage of the halo population, with sufficiently dense binning to explore interesting changes in behavior without sacrificing a sufficient population size per bin. The effect of rescaling is demonstrated in Figure \ref{fig:rescaling_example} and Figure \ref{fig:CompaSO_rescaling_example}. Self-similarity is observed as a consistency in values of $\Gamma$ over time, for a given rescaled mass bin.

\subsection{Evaluating Convergence to Self-Similarity}\label{self_similarity}
\begin{figure*}
    \centering
    \null \vspace{-5pt}
    \fig{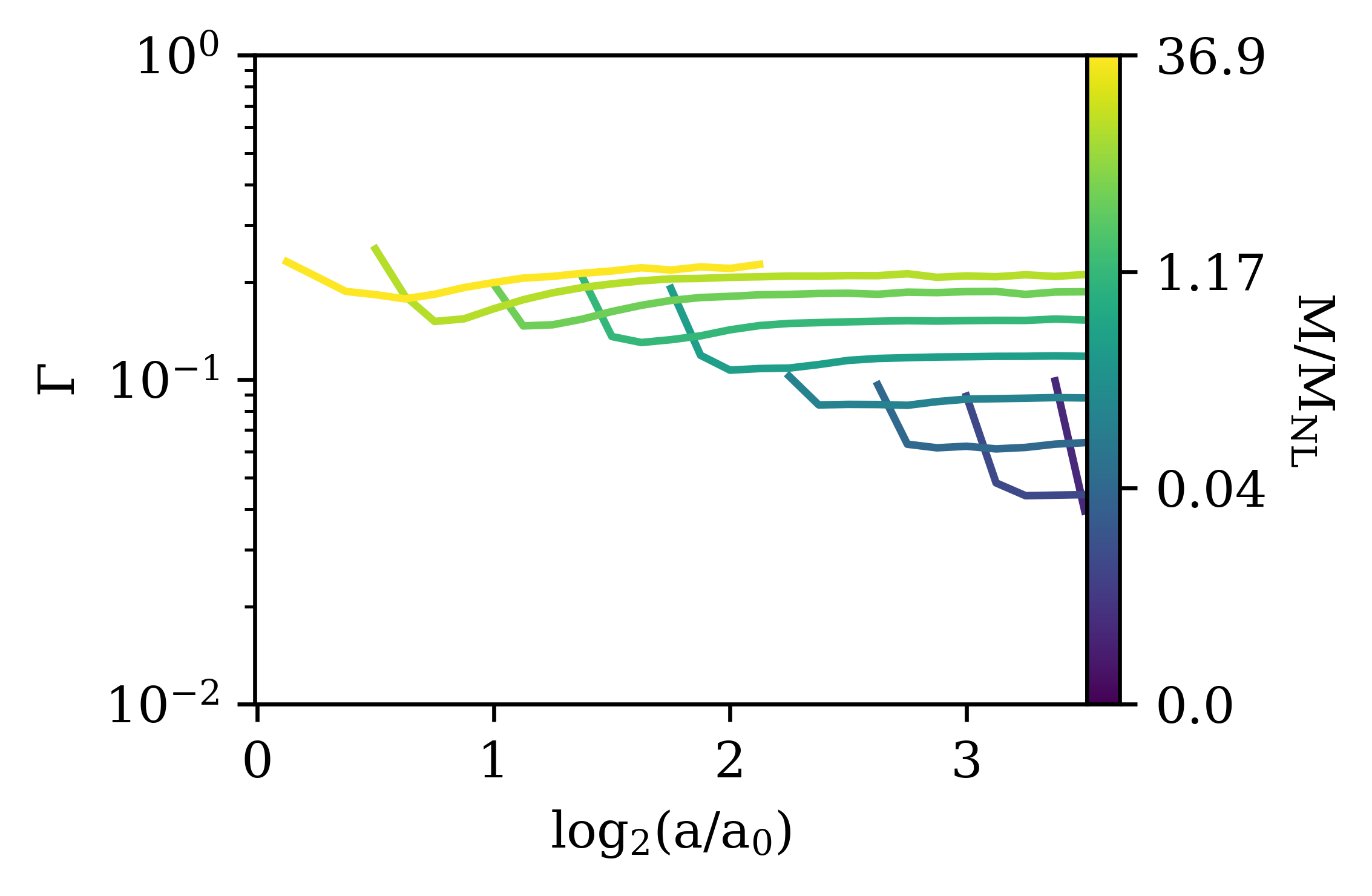}{.49\textwidth}{(a) \textsc{Rockstar}}
    \fig{Paper_Plots_gtr20/CompaSO_n15_z_mean_past_joyce_mah_M200c_gtr20_vs_a_continuous_spectrum.png}{.49\textwidth}{(b) \textsc{CompaSO}}
    \caption{\raggedright Above are the mass accretion histories ($\Gamma$) plotted against the doubling factor of the scale factor, $\log_2(a/a_0)$. Each line denotes the mass accretion history of a specific rescaled mass bin ($M/M_{\mathrm{NL}}$). In this presentation, the self-similarity discussed in section $\ref{methods}$ appears as a flattening of the $\Gamma$ vs $\log_2(a/a_0)$ line. As with Figures \ref{fig:rescaling_example} and \ref{fig:CompaSO_rescaling_example}, one can observe the very different late time behaviors of \textsc{Rockstar} (a) and \textsc{CompaSO} (b). While mass accretion histories tend towards flatness at late times for \textsc{Rockstar}, a clear drop in mass accretion history is visible for the highest rescaled mass bins for \textsc{CompaSO} results.}
    \label{fig:Rockstar_CompaSO_mah_v_a}
\end{figure*}

To evaluate the convergence of our chosen halo mass accretion metric to self-similarity we adopt the convention used in previous work \citep{Maleubre_2022,Maleubre_2023,Maleubre_2024}. In rescaled units, self-similarity is observed as the absence of temporal evolution. To quantitatively measure the convergence of halo mass accretion history to self-similarity, we then must evaluate the rescaled halo mass accretion history deviation from flatness, as a function of time (see Figure \ref{fig:Rockstar_CompaSO_mah_v_a}). Below, we provide an overview of our method.

First, we look for minimal variation over time in the mass accretion history, $\Gamma$, for each rescaled mass bin, $M/M_{\mathrm{NL}}$. Starting with the earliest times, we calculate $\Delta_1$ in regions of 5 consecutive snapshots as
\begin{equation}\label{eqn:delta1}
    \Delta_{1}\left(\frac{M}{M_{\mathrm{NL}}}, a_{i}, ..., a_{i+k}\right) = \left|\frac{\Gamma_{\mathrm{max}} - \Gamma_{\mathrm{min}}}{2\left<\Gamma\right>}\right|
\end{equation}
where the number of snapshots used, $k$ is equal to $5$ and $\Gamma_{\mathrm{max}}$ and $\Gamma_{\mathrm{min}}$ are calculated over the 5 snapshots (i.e., $a_{i}$ through $a_{i+k}$). We chose this window size conservatively to avoid spurious detection of self-similarity, while avoiding eliminating regions of substantial convergence. The general shape of the regions of convergence are robust to changes in window width, but the percent convergence of specific bins changes (see Section \ref{varying_bin_width}). If $\Delta_1 \leq X\%$, where $X\%$ is the desired level of convergence to self-similarity, we then move on to step two.

In the second step, we evaluate the convergence of each individual snapshot of $\Gamma$ for a given $M/M_{\mathrm{NL}}$ bin using $\Delta_2$ in Equation \ref{eqn:delta2}. Values are compared to the $\left<\Gamma\right>$ value of the window that passed step one. If more than one window passed step one, $\left<\Gamma\right>$ is equal to the average value of all the snapshots that passed in step one.\footnote{This is a modification of the method used in \citet{Maleubre_2024} and their earlier works, where they choose the window with the minimum $\Delta_1$ value. We find that taking the mean offers greater stability in measuring the convergence of $\Gamma$ than the previous method, however both methods yield broadly similar results.}
\begin{equation}\label{eqn:delta2}
    \Delta_2\left(\frac{M}{M_{\mathrm{NL}}}, a\right) = \left|\frac{\left<\Gamma\right> - \Gamma(\frac{M}{M_{\mathrm{NL}}}, a)}{\left<\Gamma\right>}\right|
\end{equation}
If $\Delta_2 \leq X\%$, where $X\%$ is the desired level of convergence to self-similarity, we check to see if neighboring snapshots also passed. If there exist at least three consecutive snapshots with $\Delta_2 \leq X\%$, we consider the region to be converged within $X\%$. In Appendix \ref{alternative_convergence_metrics} we explore variations of this method to demonstrate that our conclusions are robust.

\section{Numerical Simulation and Halo Finders}\label{sims_and_finders}
\subsection{\textsc{Abacus} code and Simulation 
Parameters}\label{Abacus}
\begin{figure*}
    \centering
    \null \vspace{-5pt}
    \fig{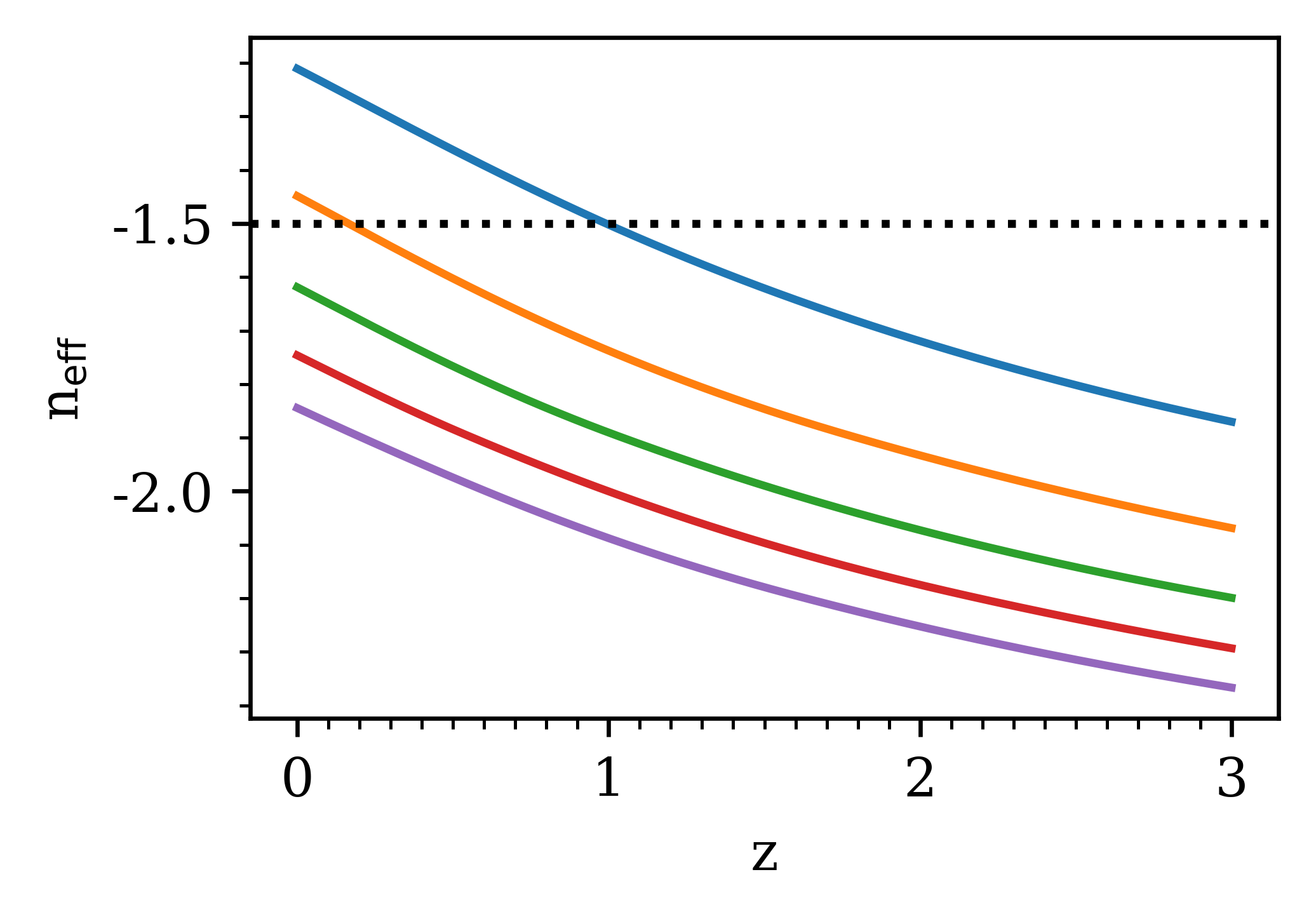}{0.49\textwidth}{(a)}
    \fig{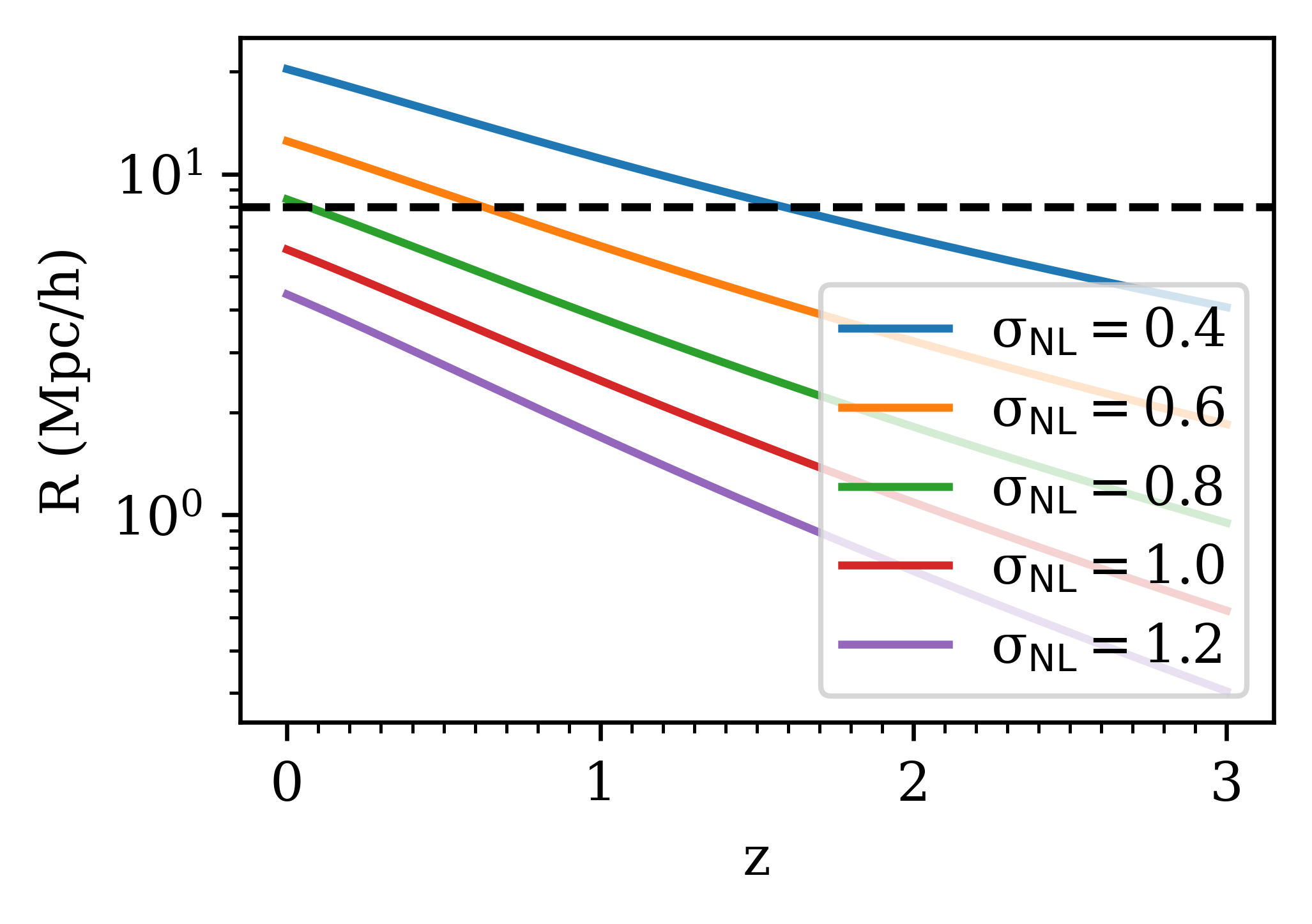}{0.49\textwidth}{(b)}
    \caption{\raggedright Plots of the effective spectral index, $n_{\mathrm{eff}}$, and scale, $R$, for a given linear top-hat variance, $\sigma_{\mathrm{lin}}\left(R, a\right) = \sigma_{\mathrm{NL}}$. Plots were made using \textsc{Colossus} \citep{Colossus}, with a \textit{Planck} 2018 cosmology \citep{Planck_2018_Cosmo}. The dotted line in plot (a) corresponds to $n_{\mathrm{eff}}=-1.5$ and the dashed line in plot (b) corresponds to $R = 8$ Mpc/$h$. The variance $\sigma_{\mathrm{NL}}$ provides a means of relating a scale and redshift in $\Lambda$CDM to a scale-free cosmology. We find that an effective spectral index of $n_{\mathrm{eff}}=-1.5$ implies a length scale of 11.15 Mpc/$h$ and a Lagrangian mass of $\sim5\times10^{14} M_{\odot}/h$, which in turn corresponds to the Lagrangian length and mass scale we expect for cluster-sized halos.}
    \label{fig:n_eff_plots}
\end{figure*}

For our simulation, we use the same $N=4096^3$ particle simulation with spectral index $n=-1.5$, generated using the \textsc{Abacus} $N$-body code \citep{Garrison_2021_Abacus}, as \citet{Maleubre_2024}. We choose a $n=-1.5$ spectral index because in $\Lambda$CDM cosmologies an effective spectral index of $n_{\mathrm{eff}} = -1.5$ corresponds to roughly the cluster scale (see Figure \ref{fig:n_eff_plots}). Given the accuracy of the $N$-body code, the high resolution of the simulation, and the simplicity of scale-free simulations versus $\Lambda$CDM simulations, the robustness limits we find can be thought of as an upper bound on other simulations for cluster scale halos.

The initial amplitude of Gaussian fluctuations was $\sigma(\ell, a_i) = 0.03$. The time step parameter is $\eta_\mathrm{acc} = 0.15$ \citep[see][for further discussion]{Joyce_2021}. The initial conditions of the simulation use \textsc{Abacus}'s second-order Lagrangian perturbation theory (2LPT) implementation and particle linear theory (PLT) corrections \citep{Garrison_2016}. Our simulation uses a spline force softening scheme \citep[see descriptions in][]{Garrison_2016} with a Plummer-equivalent length of $\ell/30$ \citep{Plummer_1911}. The impacts of the softening scheme were explored in previous works \citep[e.g.,][]{Garrison_2021_softening, Maleubre_2022}. 

We define the first output of the simulation, as in previous work, as the approximate scale factor ($a_0$) at which non-linear structures begin to form (i.e., $\sigma_{\mathrm{lin}}(\ell, a_0) = 0.56$). The subsequent simulation snapshots are spaced by a factor of $\sqrt{2}$ in the non-linear mass scale. From Equation \ref{eqn:nonlinearmass}, we find that
\begin{equation}\label{eqn:log_a_spacing}
    \Delta \log_2(a) = \frac{3 + n}{6} \Delta \log_2(M_{\mathrm{NL}}) = \frac{3+n}{12}.
\end{equation}
As a result, we can represent the time as evenly spaced points representing the doubling of the scale factor since $a_0$ using $\log_2(a/a_0)$. This has the benefit of describing the simulation in more universal terms that are then applicable to other simulations (see Section \ref{LCDM}). We examine 30 snapshots from the simulation. For a more comprehensive description of the \textsc{Abacus} $N$-body code simulation used, we refer the reader to \citet{Maleubre_2024}, and to previous works which explored the convergence of various parameter choices \citep[][]{Joyce_2021, Garrison_2021_softening, Maleubre_2022}.

\subsection{Halo Finders: \textsc{CompaSO} and \textsc{Rockstar}}\label{finders_descriptions}
In this study we test our simulation with two halo finders, the Robust Overdensity Calculation using K-Space Topologically Adaptive Refinement (\textsc{Rockstar}) \citep{Rockstar} and \textsc{CompaSO} \citep{CompaSO}. To calculate the mass accretion histories we need, we pair \textsc{Rockstar} with the merger tree algorithm \textsc{Consistent Trees} \citep{Consistent_trees} and we apply a cleaning procedure to \textsc{CompaSO} \citep{CompaSO_Merger_Trees}. In the following subsections we describe the halo finding and merger tree creation procedures of each algorithm.

\textsc{Rockstar} identifies halos with the following procedure. First, it uses a friends-of-friends algorithm \citep{FoF} to determine overdense groups of particles in the simulation volume. The normalized phase space information of each group is used to iteratively define subgroups, such that at each step 70\% of particles are linked together in subgroups. Seed halos are then placed at the deepest level of subgroup, with particles in higher subgroups assigned to the nearest halo in phase space. For an illustrated summary of this procedure, see figure 1 in \citet{Rockstar}. 

\textsc{Consistent Trees} constructs a merger tree using \textsc{Rockstar} halos, evolving the simulation backwards in time to create a consistent merger history. The initial stage does this by identifying, and sometimes creating, progenitor halos. First, it uses the positions and velocities of halos to predict their locations at the previous time step. Using that information, it then eliminates spurious progenitor-descendant relationships while identifying likely ones. Additionally, it creates progenitor halos, known as phantom halos, out of existing particles for halos with no identifiable progenitor. It also eliminates any halos without descendants that are too isolated to have been merged with another halo. 

The second stage of \textsc{Consistent Trees} focuses on eliminating bad merger trees. It removes massive halos with too high a proportion of particles coming from the aforementioned phantom halos. It also removes massive halos and subhalos that do not exist for a sufficient number of time steps. For a more in depth summary of \textsc{Consistent Trees}, see section 5 of \citet{Consistent_trees}. To construct our mass accretion histories, we use the default $M_{200c}$ mass definition from \textsc{Consistent Trees}. We identify the most massive progenitors of distinct (non-subhalo) halos. We do so by looking only at clusters with \texttt{pid} $< -1$ and then matching past halos to their descendants by comparing \texttt{desc\_id} to \texttt{id}. By selecting only unique matches or the match with most massive progenitor, we are able to construct our mass accretion histories. We limit our analysis to halos with at least 20 particles, but the phantom halos generated in \textsc{Consistent Trees} procedure go down to as low as a few particles.

\textsc{CompaSO} determines halos in three stages. Before identifying halos, a local density is estimated at each particle location using a kernel width of $0.4\ell$, where $\ell$ is again the initial particle spacing. Particles with a sufficiently high local density are grouped into what are called L0 groups, using a friends-of-friends algorithm. Within each L0 group, a set of halos, L1 halos, are calculated. The particle with the peak local density inside an L0 group is treated as the nucleus of an L1 halo. Particles are tentatively assigned to the L1 halo if they fall within a radius determined with a threshold density. This L1 identification process is then repeated using all particles that fall outside 80\% of the L1 radius. In this case the nucleus of any additional L1 halos is required to have a local density greater than all particles within some predefined radius. Particles are assigned to these additional L1 halos if they were previous unassigned or if their enclosed density with respect to the new nucleus is twice that of the their enclosed density with respect to the nucleus of the previously assigned halo. This process is repeated until a minimum density threshold is reached. Finally, subhalos, called L2 halos, are identified for each L1 halo. They are assigned using the L1 halo identification method, but within a given L1 halo. Note that \textsc{CompaSO} imposes a strict $30$ particle cutoff for halos.

Merger trees are constructed and cleaned using the procedure described in \citet{CompaSO_Merger_Trees}. The algorithm does so by tracing a subsample of 10\% of the particles in a halo across time. Starting with some L1 halo, the algorithm looks back one previous time step and identifies all L1 halos that existed within $4$ Mpc. Using this list of halos, a list of candidate progenitors is made by evaluating which historical halos have a non-zero fraction of particles from the current halo. If the fraction of particles in the candidate shared between the candidate and the current halo is above the threshold fraction, the candidate is considered a progenitor. The progenitor halo that contributes the most particles to the current halo is called the main progenitor. This procedure is completed for halos two time steps back in time from the current time, but only the main progenitor is recorded, and is referred to as the main progenitor preceding. This procedure is repeated for all halos at all times where a sufficient number of time steps are available. There is an additional cleaning procedure, analogous to the second stage of \textsc{Consistent Trees}. For a given halo at time $z_i$, the algorithm identifies a redshift, $z_{\mathrm{max}}$, at which the halo on the main progenitor branch of the current halo reaches its greatest mass. If this greatest mass is sufficiently larger than the mass of the current halo, the current halo is flagged for cleaning. The flagged halo is merged with a contemporaneous neighboring halo that is the most massive descendant of the maximum mass main progenitor identified at $z_{\mathrm{max}}$. The combined halo then remains combined for all future time steps. The effect of this procedure is to reduce the breaking up of larger halos inflicted by the stricter spherical over density cuts imposed by the primary \textsc{CompaSO} procedure. \citet{Maleubre_2024} demonstrated that the cleaning procedure improved convergence to self-similarity in the halo mass function. We do not directly test the impact of omitting the cleaning procedure in this work, however in Section \ref{compaso_results} we discuss the ways in which differences in \textsc{CompaSO} and its cleaning procedure as compared to \textsc{Rockstar} and \textsc{Consistent Trees} might impact the preservation of self-similarity.

With \textsc{Rockstar} we are able to retrieve halo data for all 30 simulation snapshots, and therefore can analyze the mean mass accretion history information of 29 snapshots. When using halo data from \textsc{CompaSO}, due to an issue with the data, we are limited to 27 snapshots of halos, and therefore 26 snapshots of mass accretion history information. 

\section{Results}\label{results}
\subsection{\textsc{Rockstar}}\label{rockstar_results}

\begin{figure*}
    \centering
    \null \vspace{-5pt}
    \fig{Paper_Plots_gtr20/Rockstar15_mean_past_joyce_mah_M200c_gtr20_FW_WW5_M_M_NL_vs_a_convergence_MT2_UVM_min_1000.png}{0.49\textwidth}{(a) \textsc{Rockstar} Convergence vs Rescaled Mass}
    \fig{Paper_Plots_gtr20/Rockstar_n15_mean_past_joyce_mah_M200c_gtr20_FW_WW5_N_vs_a_convergence_MT2_UVM_min_1000.png}{0.49\textwidth}{(b) \textsc{Rockstar} Convergence vs Halo Particle Count}
    \caption{\raggedright Plots of the convergence of mass accretion histories, calculated using the \textsc{Rockstar} halo finder are shown. In plot (a) the convergence is plotted as a function of the rescaled mass bin ($M/M_{\mathrm{NL}}$) and the doubling factor of the scale factor since the first halos collapsed ($\log_{2} \left(a/a_0\right)$). Plot (b) shows the same convergence, but replaces the rescaled mass bin with the particle count per halo ($N$). This transformation is possible using the second definition of $M_{\mathrm{NL}}$ in Equation \ref{eqn:nonlinearmass}. From plot (b), one can get a sense of the bounds of convergence. Convergence is strongest when $\log_2(a/a_0) \geq 2$. Within that time frame, the converged particle count varies from $\sim  20$ to $\sim 2\times10^4$. Centered in this area, there is a concave region of convergence, with a continuous region of high (1\%) convergence that is surrounded by smaller shells of lesser convergence. The very upper bounds of convergence, in terms of particle count, are near $10^5$. This upper bound of convergence at late times appears when the halo population limit ($> 1000$ halos) is reached, so it is probable that the limit is imposed by the finite volume of the simulation, and not the halo finder. The lower bound of convergence demonstrates interesting temporal behavior. At very early times, $\log_2 (a/a_0) \sim 0.5$, 2\% convergence is observed. Immediately following this the lower bound of convergence is pushed up to $N\sim10^2$. Convergence then begins to reach lower particle counts, before reaching the minimum lower bound of $N=20$ at $\log_2 (a/a_0) = 3$. Beyond this point, our choice of convergence window width begins to influence our ability to measure convergence, and no further convergence is measured. A full analysis of this behavior can be found in Section \ref{rockstar_results}.}
    \label{fig:Rockstar_convergence}
\end{figure*}

Figure \ref{fig:Rockstar_convergence} displays the results of our convergence procedure using halos found with \textsc{Rockstar}. The results demonstrate a remarkable range of robustness for our definition of mass accretion history, with halos ranging from a few tens of particles to nearly 100,000 particles demonstrating 2\% convergence to self-similarity. The region of highest convergence is found at late times ($\leq 1$\% at $\log_2(a/a_0) \geq 2$). At its greatest extent, when the scale factor is roughly eight times the scale factor at non-linear collapse ($a_0$), the region of high convergence stretches almost continuously from as low as a couple hundred particles to as high as $\sim 20,000$ particles. 

It is unsurprising that the most massive halos ($\gtrsim1000$ particles) demonstrate greater convergence at later times. The hierarchical nature of structure formation means that larger structures take longer to form. As a consequence, as the simulation evolves, the number of very massive halos grows. Since we require there to be more than 1000 halos for a region to be probed, an increase in the number of very massive halos at later times allows new regions of parameter space to be probed. The depth of convergence for more massive halos is similarly improved by their increased number. The mass accretion history we use is a mean of a distribution. When the population of halos is low, that distribution, which can be thought of as a random sample of an underlying distribution of possible mass accretion histories for halos of that rescaled mass bin, will be noisier. This noisiness will prevent higher levels of convergence. However, as the population of very massive halos grows, the noisiness of the sample of the mass accretion history distribution decreases, and thus convergence to self-similarity is more likely. Therefore, when using \textsc{Rockstar}, the convergence limits of the most massive halos are dictated by the finite box size of the simulation, not the properties of the halo finder itself. Finally, it is worth noting that halo masses in this simulation are discrete. More massive halos therefore have an advantage in achieving convergence over less massive halos, as their increased masses mean that the integer mass of halos has a less prominent effect on the possible mass accretion histories available to a halo.

For intermediate masses ($\sim100-1000$), similar reasoning explains the improvement in convergence over time. At very early times intermediate mass halos are drawn from the exponential tail of the halo mass function. In this region, there are very few halos of similar mass. This sparsity makes the mean mass accretion history noisy. At later times intermediate mass halos are drawn from the power-law regime of the halo mass function, meaning there are many halos of similar mass, thus stabilizing the mean mass accretion history. At early times the intermediate mass halos are some of the most massive halos in the simulation and therefore are relatively isolated. At late times there are large populations of small ($<100$) and large ($>1000$) halos. This means that for all intermediate mass halos there is a sufficiently large population of halos to accrete mass from, and to be accreted on to, to avoid being dominated by numerical effects.

\begin{figure*}
    \centering
    \null \vspace{-5pt}
    \gridline{\fig{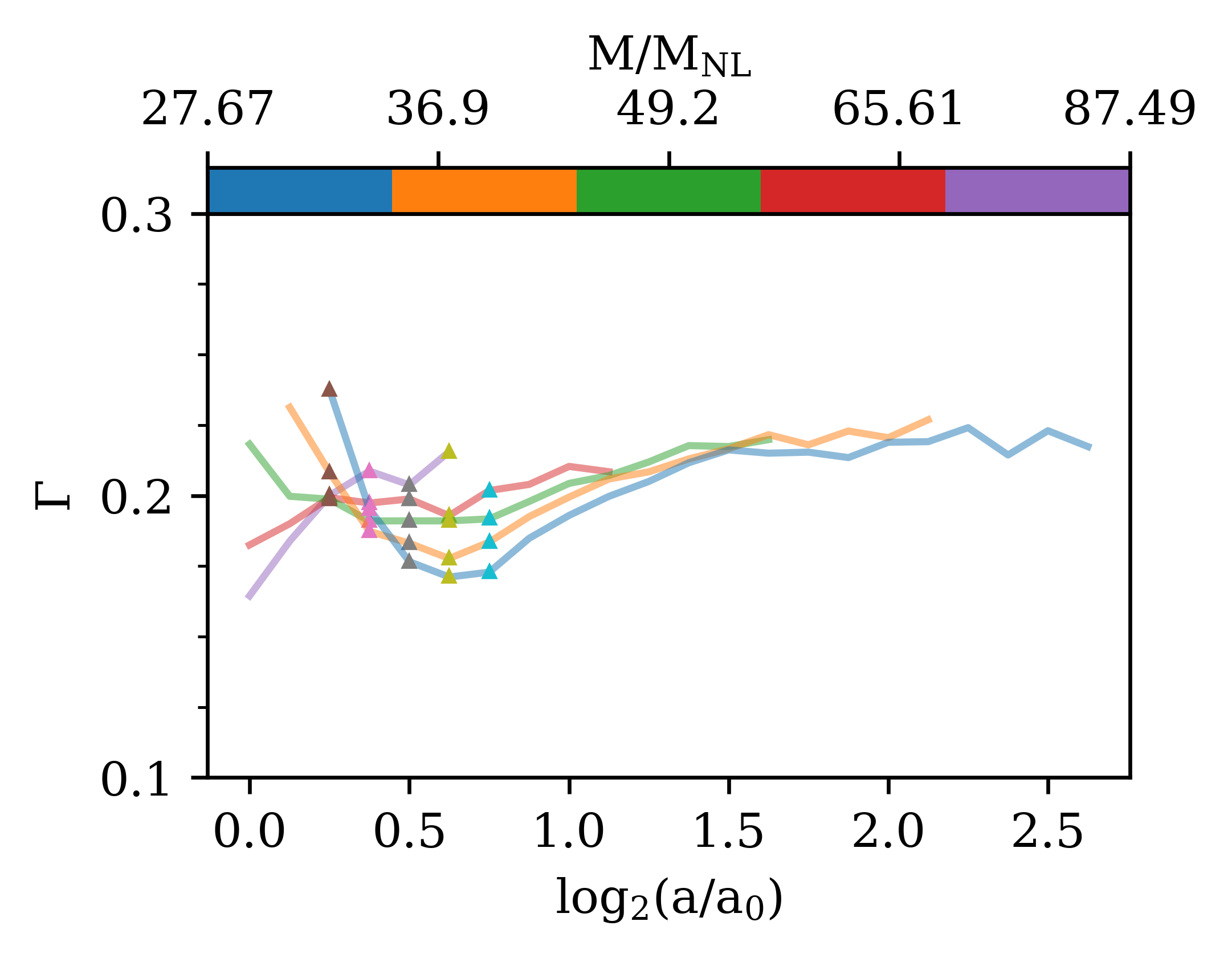}{0.49\textwidth}{(a) Mean mass accretion histories of 5 high $M/M_{NL}$ bins}
    \fig{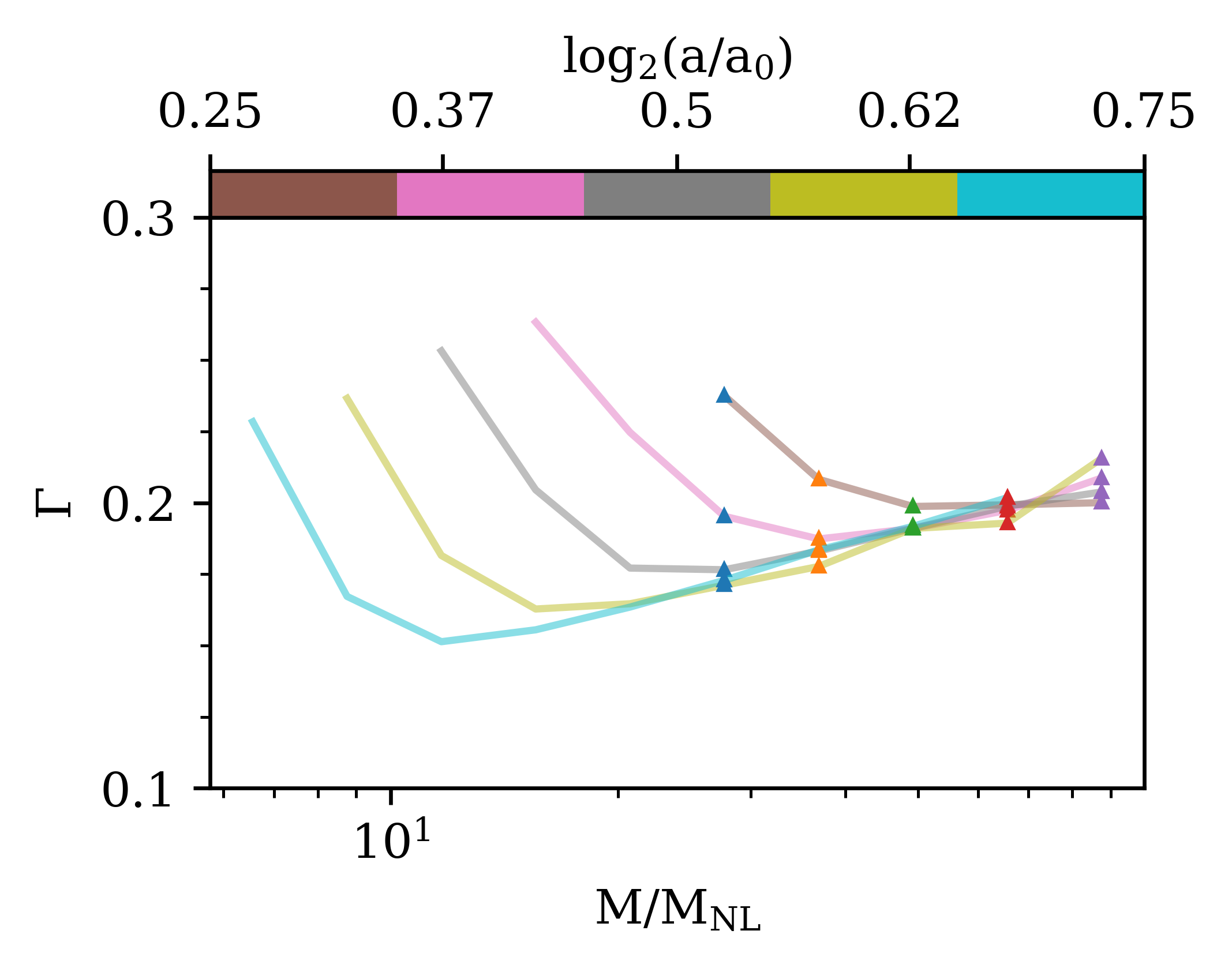}{0.49\textwidth}{(b) Mean mass accretion history of 5 early snapshots}}

    \caption{\raggedright  Above are plots illustrating the mass accretion histories of low mass halos in the early time 2\% convergence region for halos found using \textsc{Rockstar}. Plot (a), in the style of Figure \ref{fig:Rockstar_CompaSO_mah_v_a}, shows the mean mass accretion history ($\Gamma$) of five rescaled mass bins, as a function of time (in units of the doubling of the scale factor since $a_0$). The triangles indicate the 5 times probed. Plot (b), like Figures \ref{fig:rescaling_example} and \ref{fig:CompaSO_rescaling_example}, shows the mean mass accretion history of the 5 snapshots as a function of rescaled mass bin. Here the triangles indicate the 5 rescaled mass bins probed. The chosen combination of rescaled mass bins and snapshots corresponds to the top left corner of Figure \ref{fig:Rockstar_convergence} (a), which in turn corresponds to the bottom left corner of plot (b) of the same figure. In plot (a) we can see the early region of flatness that drives the 2\% convergence on the green line. We can also see how that flatness appears out of a smooth transition in the shape of the $\Gamma$ curve. For the highest $M/M_{NL}$ bin shown (purple), $\Gamma$ starts lower and transitions to higher values at later times. As the $M/M_{NL}$ bin decreases, what was once a bump (purple) in $\Gamma$ at $log_2 (a/a_0)\sim0.5$ flattens (green) before becoming a dip (blue). We can also see that a new, later region of flatness appears near $log_2 (a/a_0)\sim1.5$. Taken together, we can see that there is a brief region of apparent self-similarity at early times. This region disappears as a later region of self-similarity appears with a different mean mass accretion history value. Since the early region of apparent self-similarity is transitory, and, more importantly, converges to a different value than later times, we view it as dubious. We discuss a possible reason for this pseudo-convergence in Section \ref{rockstar_results}.}
    \label{fig:Rockstar_Transient_Convergence}
\end{figure*}

Beyond illustrating the simulation and halo finder's deep fidelity to self-similarity, Figure \ref{fig:Rockstar_convergence} also reveals interesting convergence behavior for low mass halos. At very early times, there is an isolated region of 2\% convergence, that then disappears entirely. This behavior is explored further in Figure \ref{fig:Rockstar_Transient_Convergence}. The evidence suggests that the sudden appearance and disappearance of 2\% convergence is an artifact of the choice convergence window width (see Fig.~\ref{fig:rockstar_vary_window_N_v_a_convergence}.) However, there is some level of real convergence to self-similarity underlying that behavior. At early times, low particle count halos of the same rescaled mass bin fall within a tight range of mean mass accretion histories, demonstrating the flatness behavior discussed in Section \ref{self_similarity}. To understand this, one must consider the behavior holistically. The long term trend of mean mass accretion histories, for all evaluated rescaled mass bins and times, is to decrease as the simulation evolves forward in time. At very early times, the mass accretion history distribution is dominated by very massive accretors and comparatively fewer negative accretors. As the simulation evolves, this balances shifts, with comparatively fewer extreme positive accretors, more negative accretors, and many more small but positive accretors. It appears that for relatively small halos at early times, the change in extreme positive accretors vs negative accretors is sufficiently balanced to stabilize the mean and create the appearance of convergence. Given the uncertainty around this behavior, we consider this region of convergence to be dubious.

Convergence at very low particle counts is not observed again until much later times, by which point the scale factor has quadrupled. This convergence develops gradually and continuously, starting at higher particle count halos at earlier times and gradually reaching very low mass halos at later times.  It is possible that the softening length, which gets smaller in comoving coordinates as the simulation evolves, is improving convergence \citep{Garrison_2021_softening}. Similarly, as the simulation evolves with time, it may begin to lose its memory of the initial lattice of particles \citep[see][for further discussion]{Joyce_2021, Maleubre_2022}. At very late times ($\log_2(a/a_0)\geq3$), we are limited not by the simulation or the halo finder, but by the convergence estimation procedure itself. As seen most clearly in Figure \ref{fig:Rockstar_convergence} (a), for the lowest mass halos at late times, there are less than five time steps available at a given rescaled mass bin. Since our procedure limits us to regions of stability at least five snapshots wide, we are unable to probe these regions. Alternative convergence metrics, explored in Appendix \ref{alternative_convergence_metrics}, imply that these unexplored regions also are converged.

\subsection{\textsc{CompaSO}}\label{compaso_results}
Figure \ref{fig:CompaSO_convergence} illustrates our results for halos found using \textsc{CompaSO}. We can see that there is a region of high ($\leq2$\%) convergence at late times ($\log_2 (a/a_0) > 2$), ranging from several hundred to a few thousand particles per halo. Like with the \textsc{Rockstar} results, this late time convergence is likely driven by sufficient evolution away from the non-physical length scales introduced in the initial conditions of the simulation.

The bounds of convergence, both of the $\leq2$\% region, and of the much larger $\leq10$\% region, are driven by the choice of halo finder. \textsc{CompaSO} imposes a strict spherical overdensity definition for identifying halos and their constituent particles. The \textsc{CompaSO} cleaning procedure, as described in Section \ref{finders_descriptions}, attempts to correct this behavior by re-identifying some small neighboring halos and flyby halos as being part of their larger neighbors. This cleaning procedure, while needed, is insufficient to preserve the robustness of the mass accretion history of very large halos. The upper bound of convergence in Figure \ref{fig:CompaSO_convergence} demonstrates a nearly flat, in particle count per halo space, cutoff for high mass halos. Beyond this point halos are too large, and too diffuse at high radii, and therefore inappropriately broken up by \textsc{CompaSO}'s halo identification procedure. Similarly, it is possible that the cleaning procedure, in re-identifying some small halos as being simply subhalos of a larger halo, is incorrectly destroying what should be independent halos. This could also be imposing the bound on convergence for low mass halos. In the case of the low mass halos, another important barrier to low mass convergence is the 30 particle limit for halos, which is 10 particles greater than the limit we impose on the \textsc{Rockstar} results.

Despite these limitations, the \textsc{CompaSO} results still demonstrate convergence over an impressive range in time and particle count. The convergence of sub-100 particle halos is consistent from $a\sim \sqrt{2}a_0$ to very late times. The loss of convergence for very late times above this $\sim 50$ particle boundary is, like in the \textsc{Rockstar} case, driven by our choice of window width in determining convergence (as is evident in Figures \ref{fig:compaso_vary_window_N_v_a_convergence} and \ref{fig:tapered_tail_N_v_a_convergence} in appendix \ref{alternative_convergence_metrics}). Similarly, when not limited by the number of available high mass halos, there is a stable boundary of convergence at $\sim5000$ particles per halo. This convergence upper bound is similar to that found in \citet{Maleubre_2024}.

\begin{figure*}
    \centering
    \null \vspace{-5pt}
    \fig{Paper_Plots_gtr20/CompaSO15_mean_past_joyce_mah_M200c_gtr20_FW_WW5_M_M_NL_vs_a_convergence_MT2_UVM_min_1000.png}{0.49\textwidth}{(a) \textsc{CompaSO} Convergence vs Rescaled Mass}
    \fig{Paper_Plots_gtr20/CompaSO_n15_mean_past_joyce_mah_M200c_gtr20_FW_WW5_N_vs_a_convergence_MT2_UVM_min_1000.png}{0.49\textwidth}{(b) \textsc{CompaSO} Convergence vs halo particle count}
    \caption{\raggedright Above are plots similar to those shown in Figure \ref{fig:Rockstar_convergence}, but for mass accretion histories obtained using the \textsc{CompaSO} halo finder. A region of high convergence, albeit lower than that seen in Figure \ref{fig:Rockstar_convergence}, is visible for $\log_2(a/a_0) > 2$. The range of particle counts at which 2\% convergence achieved is approximately $5\times10^2$ to $6\times10^3$. The small size of this region is partially imposed by a clear upper bound in convergence at $N\sim10^4$, well below the limit imposed by the limited very high mass halo population of the simulation. The lower bound of convergence does not display as obvious a time evolution as seen in Figure \ref{fig:Rockstar_convergence}. Instead, there is a mostly flat lower bound of 10\% convergence at $N\sim50-60$. The 5\% convergence does display time evolution, however, with convergence reaching lower particle counts as $\log_2(a/a_0)$ goes from $1$ to $\gtrsim2$. Convergence also appears to weaken past $\log_2(a/a_0)\gtrsim3$. Further analysis can be found in Section \ref{compaso_results} and Section \ref{comparing_finders}.}
    \label{fig:CompaSO_convergence}
\end{figure*}

\section{Discussion}\label{discussion}
\subsection{Comparing Halo Finders}\label{comparing_finders}
\begin{figure*}
    \centering
    \null \vspace{-5pt}
    \fig{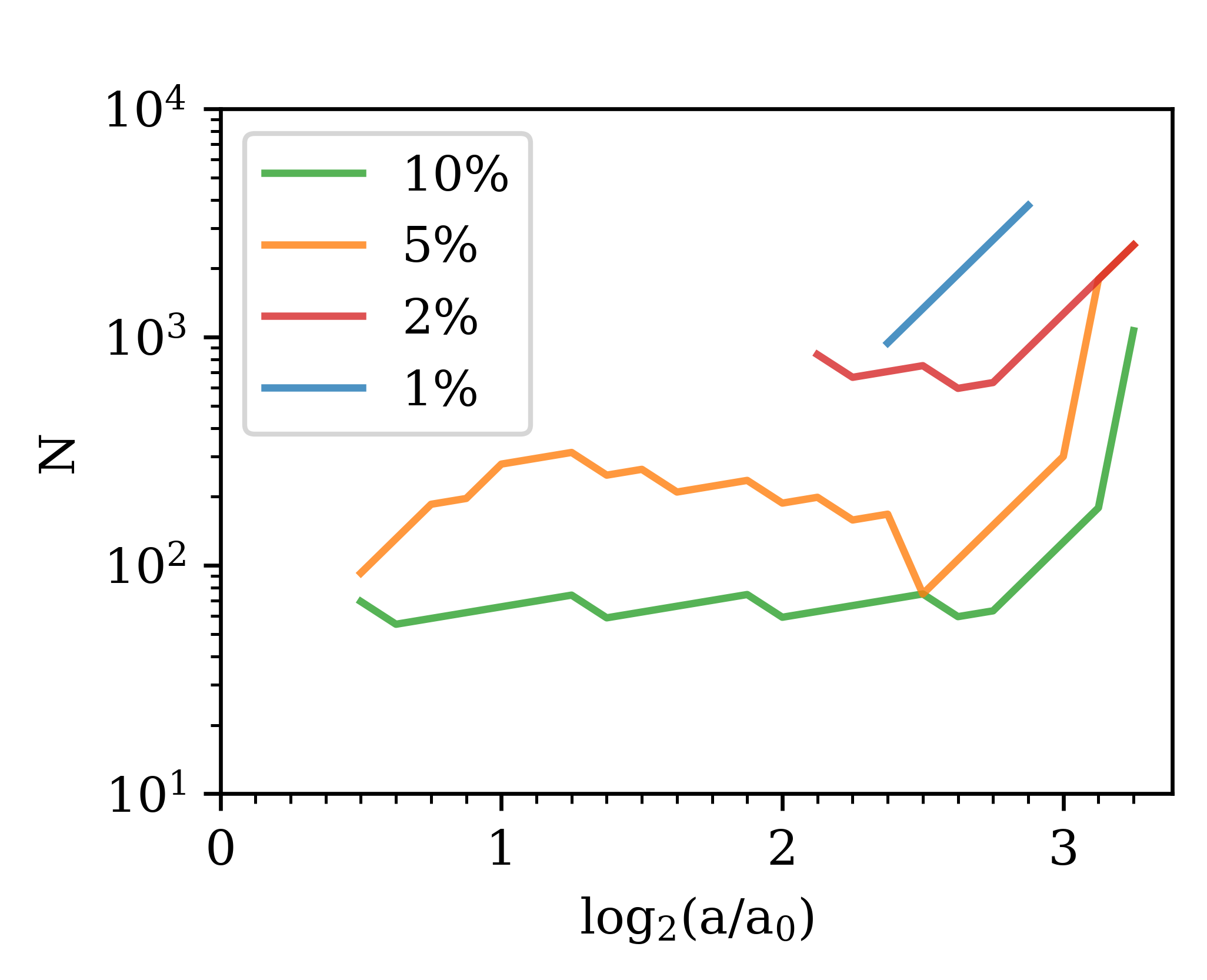}{0.49\textwidth}{(a) \textsc{CompaSO}}
    \fig{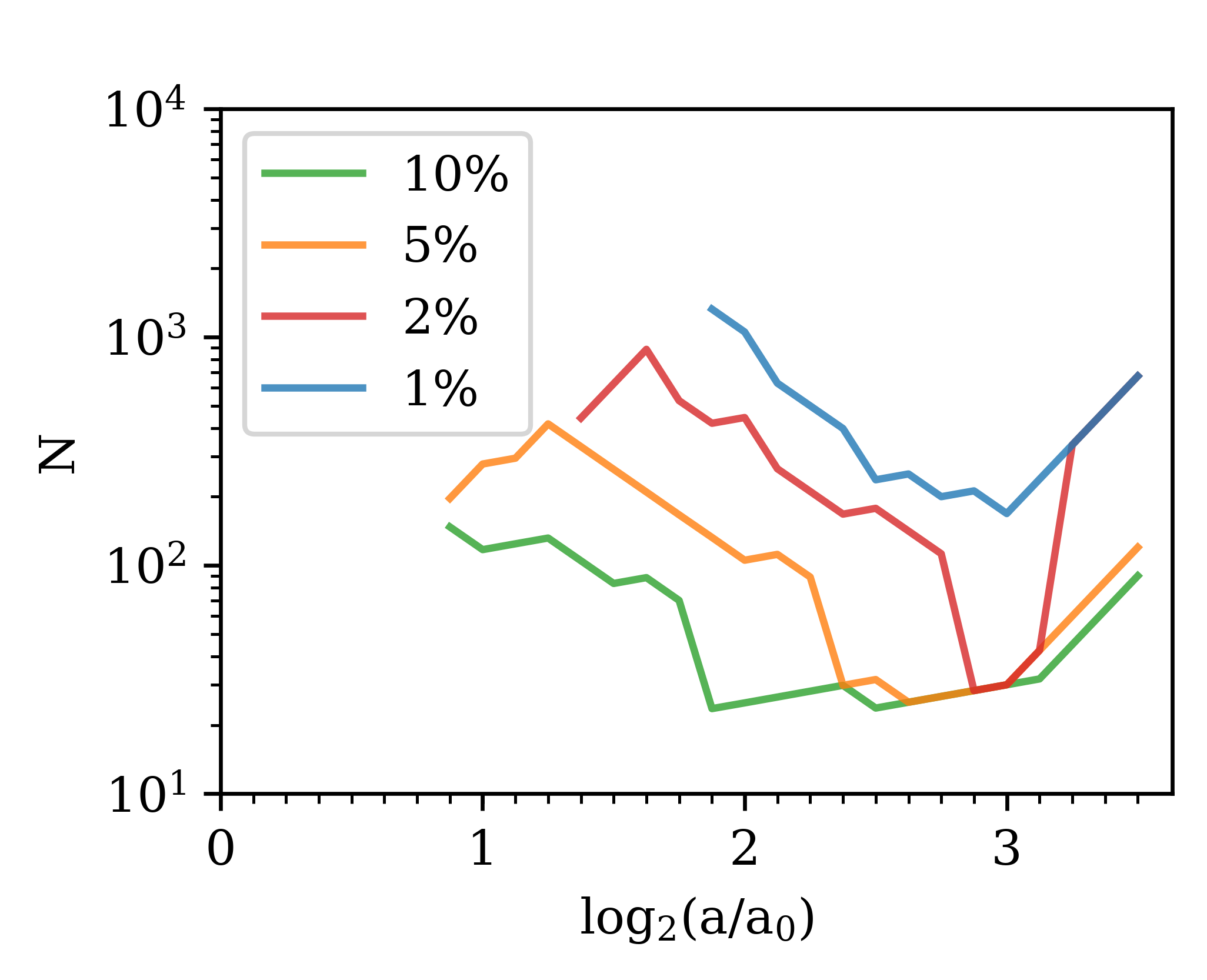}{0.49\textwidth}{(b) \textsc{Rockstar}}
    \caption{\raggedright Above are plots of the \textsc{CompaSO} and \textsc{Rockstar} convergence lower bounds. For a given snapshot, the convergence lower bound is defined as the lowest particle mass at which $X$\% of convergence is achieved. Furthermore, we require there to be some continuity, meaning that more massive halos within the snapshot must also be converged to the same level. Finally, we remove the very early time convergence salient for \textsc{Rockstar}, for the reasons discussed in Section \ref{rockstar_results} and Figure \ref{fig:Rockstar_Transient_Convergence}. The lower bound plots illustrate that \textsc{Rockstar} obtains much better convergence at later times. We also see that \textsc{CompaSO} does not demonstrate the same intensity in the evolution of convergence for low mass halos. Instead we see a modest time evolution in the 5\% convergence and no time evolution for 10\% convergence. We discuss why \textsc{CompaSO} has weaker low mass convergence in Section \ref{compaso_results}.}
    \label{fig:lower_bounds}
\end{figure*}

Like in \citet{Maleubre_2024}, we find that \textsc{Rockstar} generally preserves self-similarity better than \textsc{CompaSO}. While the results for both halo finders show a region of high convergence in roughly the same particle count versus scale factor size space, the depth and breadth of convergence for \textsc{Rockstar} is greater. \textsc{Rockstar} also demonstrates a superior convergence for very high mass halos at late times, for the reasons discussed in Section \ref{compaso_results}. \textsc{Rockstar} also has superior convergence for low mass halos at late times, as seen in Figure \ref{fig:lower_bounds}.

While \textsc{Rockstar} has a number of advantages with regards to mass accretion history fidelity, \textsc{CompaSO} has an absolute advantage in terms of computational cost and speed. Given this difference in computational cost, it is not difficult to imagine scenario's were \textsc{CompaSO}'s sub-optimal convergence is preferable. In said cases, our results suggest that \textsc{CompaSO}'s mass accretion histories are safest when used for sizeable late time halos. Less conservatively, halos with between a few thousand and roughly one hundred particles are converged. 

It is important to note that we cannot make absolute claims as to which halo finder is more physically accurate. This analysis, in determining deviations from self-similarity, is able to determine the resolutions at which the halo finder is no longer reliable. Testing the physical accuracy of the halo finders, or the simulation itself, requires a combination of scale-free validation, non-scale-free validation, and comparing simulated observables to real observables.

\subsection{Applications to Non Scale-Free Cosmologies}\label{LCDM}
While the convergence bounds provided by this work were found using only a single simulation and two halo finders, the results can be applied more broadly. The redshift dependent results of this work are presented as a function of $a/a_0$. Given the initial particle spacing $\ell$, one can use our definition of $a_0$ (see Section \ref{Abacus}) to determine $a_0$ for any other simulation. With this information, we can apply our results to $\Lambda$CDM simulations on sufficiently large scales such that the effective spectral index is $n_{\mathrm{eff}}\sim-1.5$. As discussed in Section \ref{Abacus}, this means our results are already applicable as an upper bound to galaxy cluster scales in other simulations. To apply our results more broadly still, we would need to explore the dependence of mean mass accretion history convergence as a function of the initial power spectrum. Preliminary work suggests that changing the spectral index has a real but minor impact on the convergence limits of the mean mass accretion history, but further work is needed. Comparing $n=-1.5$ and $n=-2.0$ results from \citet{Maleubre_2024} using the halo mass function also suggest that difference is minor. Therefore it is possible that our results are not strongly dependent on the initial conditions, and apply more generally to $\Lambda$CDM simulations.

A second possible caveat to this is that our results are derived from an EdS cosmology, thus one should be more cautious in applying our results to $\Lambda$CDM simulations near $z=0$, where deviations from EdS become more significant. Work by \citet{Joyce_2021} on the particle two-point correlation function (2PCF), suggested that the convergence of the 2PCF at small scales and late times might be cosmological model independent, and thus applicable to both a scale-free EdS simulation and a $\Lambda$CDM simulation at late times. In brief, the resolution of the 2PCF at small scales at late times is dominated by two-body scattering. This interaction is independent of the global properties of the simulation, and therefore cosmological model independent. Applying these conclusions to our work with the mean mass accretion history is less straightforward. The mean mass accretion history is by definition dependent on inter-halo interactions, and therefore the halo mass function, which is more sensitive to the choice of cosmological model. Work done by \citet{Maleubre_2024} using the halo-center 2PCF, which in principle would also depend on inter-halo interactions, suggests that the dependence on the cosmological model is weak, however. This suggests that our late time low mass results may indeed be nearly model independent, and therefore applicable even to low-redshift $\Lambda$CDM simulations.

\section{Conclusion}\label{conclusion}
In this work we probe the robustness of the halo mass accretion histories in a scale-free simulation generated using the \textsc{Abacus} $N$-body code. We do so by taking advantage of the expected self-similarity of dimensionless properties rescaled by the only physical scale in the simulation, the scale of non-linearity. We test the added effects on mass accretion robustness from the choice of halo finder, examining two different halo finders, \textsc{Rockstar} and \textsc{CompaSO}. We find that \textsc{Rockstar} demonstrates superior convergence than \textsc{CompaSO}. Both simulations demonstrate a wide range of convergence, with near 5\% convergence for the duration of the simulation for halos with particles counts as low as $\sim100$. This lower bound in particle count reduces further at late times, with an absolute minimum of $\sim70$ and $\sim30$ particles per halo demonstrating 5\% convergence for \textsc{CompaSO} and 2\% convergence for \textsc{Rockstar}, respectively. The upper bound particle limit also evolves with the simulation, reaching a maximum of $\sim10^4$ particles per halo when using \textsc{CompaSO} and maximum of $\sim10^5$ when using \textsc{Rockstar}. In \textsc{CompaSO} this upper limit appears to be a product of the limitations of the halo finder, but for \textsc{Rockstar} this upper bound occurs because the number of halos falls too low, suggesting we are limited by the finite volume of the simulation.

Halo mass accretion history is a key property for understanding galaxy clusters, and for using galaxy clusters to constrain cosmology. Much of the research into galaxy clusters and mass accretion history is reliant on cosmological simulations, and therefore in need of verification methods like the one discussed in this work. While the results of this work are easily applied to investigations of mass accretion history that use the \textsc{Abacus} $N$-body code, such as Warburton et al (in prep), the method used can be applied more broadly. Any code capable of creating a dark matter only simulation can be used to construct a scale-free simulation, which can then be probed using our method. Moreover, because \textsc{Abacus} is a highly accurate N-body code, the results here can be taken as the upper limit on possible convergence for other $N$-body codes, at least when using similar halo finding methods. Similarly, by taking advantage of our choice of spectral index, the results can provide limits of convergence on $\Lambda$CDM simulations at the galaxy cluster scale. By varying meta-parameters, like resolution and box size, as well as cosmological initial conditions, like spectral index, one can apply the convergence limits learned via this method to a variety of other cosmologies and scales. Our self-similarity test can also be applied to any other scale-free simulation and on other dimensionless clustering properties within a scale-free simulation. Doing so is essential to verifying the accuracy of state-of-the-art simulations, and therefore essential for verifying the inferences dependent on those simulations.

\begin{acknowledgements}

We thank Michael Joyce, Sara Maleubre Molinero, Sownak Bose, and Daniel Eisenstein for comments and suggestions for this work. Previous work on scale free simulations by them, like \citet{Maleubre_2024}, was also integral to this work. We also thank the Flatiron Institute Center for Computational Astrophysics Pre-Doctoral Program, which allowed for this collaboration.

The material presented is based upon work supported by NASA under award No. 80NSSC22K0821.

The simulations in this work used resources of the Oak Ridge Leadership Computing Facility (OLCF), which is a DOE Office of Science User Facility supported under Contract DE-AC05-00OR22725. The work was conducted under OLCF projects AST135 and AST145, the latter through the Department of Energy ALCC program. Most of the analysis was performed on computational resources supported by the Scientific Computing Core at Flatiron Institute, a division of Simons Foundation.
\end{acknowledgements}

\bibliography{myref.bib}

\appendix
\section{Alternative Convergence Metrics}\label{alternative_convergence_metrics}
\subsection{Varying Convergence Window Widths}\label{varying_bin_width}

\begin{figure*}
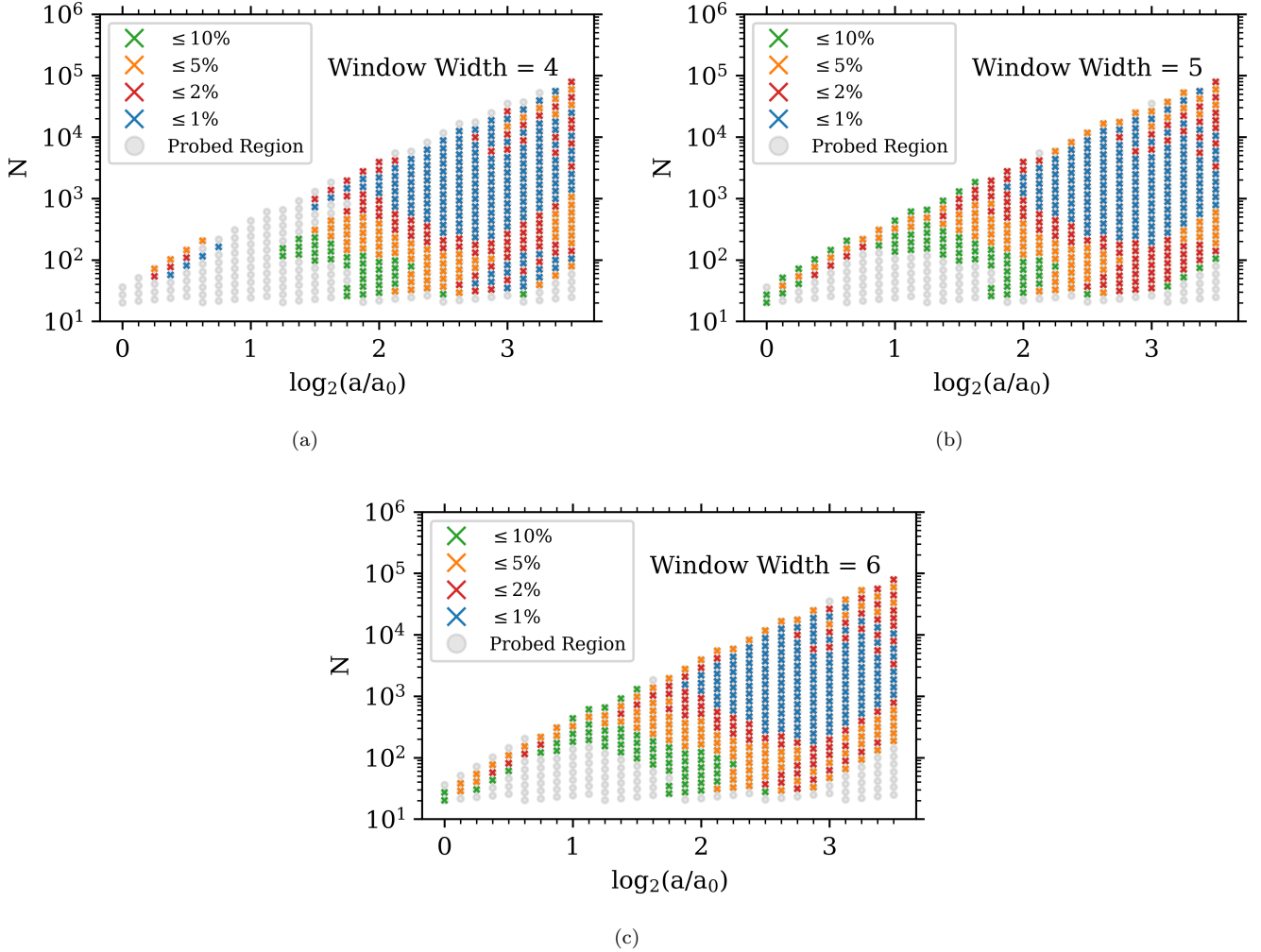

    \centering
    \null \vspace{-5pt}
    \gridline{\fig{Appendix_Plots_gtr20/Rockstar_n15_mean_past_joyce_mah_M200c_gtr20_FW_WW4_N_vs_a_convergence_MT2_UVM_with_text_min_1000.png}{0.49\textwidth}{(a)}
    \fig{Appendix_Plots_gtr20/Rockstar_n15_mean_past_joyce_mah_M200c_gtr20_FW_WW5_N_vs_a_convergence_MT2_UVM_with_text_min_1000.png}{0.49\textwidth}{(b)}}
    \gridline{\fig{Appendix_Plots_gtr20/Rockstar_n15_mean_past_joyce_mah_M200c_gtr20_FW_WW6_N_vs_a_convergence_MT2_UVM_with_text_min_1000.png}{0.49\textwidth}{(c)}}
    \caption{\raggedright Plots of the \textsc{Rockstar} convergence for different choices of window width in Equation \ref{eqn:delta1}. As discussed in Appendix \ref{varying_bin_width} and Section \ref{rockstar_results}, early time convergence is sensitive to changes in window width. The convergence regions produced using a four snapshot wide window are the noisiest, especially at higher particle counts. Here we suspect that a combination of lower halo counts and a heightened sensitivity to noise in the $\left<\Gamma\right>$ vs time curve are to blame. However, the differences between five and six snapshot wide widths is minimal, suggesting that some stability is achieved by five snapshots. In all cases the absolute lower and upper bounds of convergence remain the same.}
    \label{fig:rockstar_vary_window_N_v_a_convergence}
\end{figure*}

\begin{figure*}
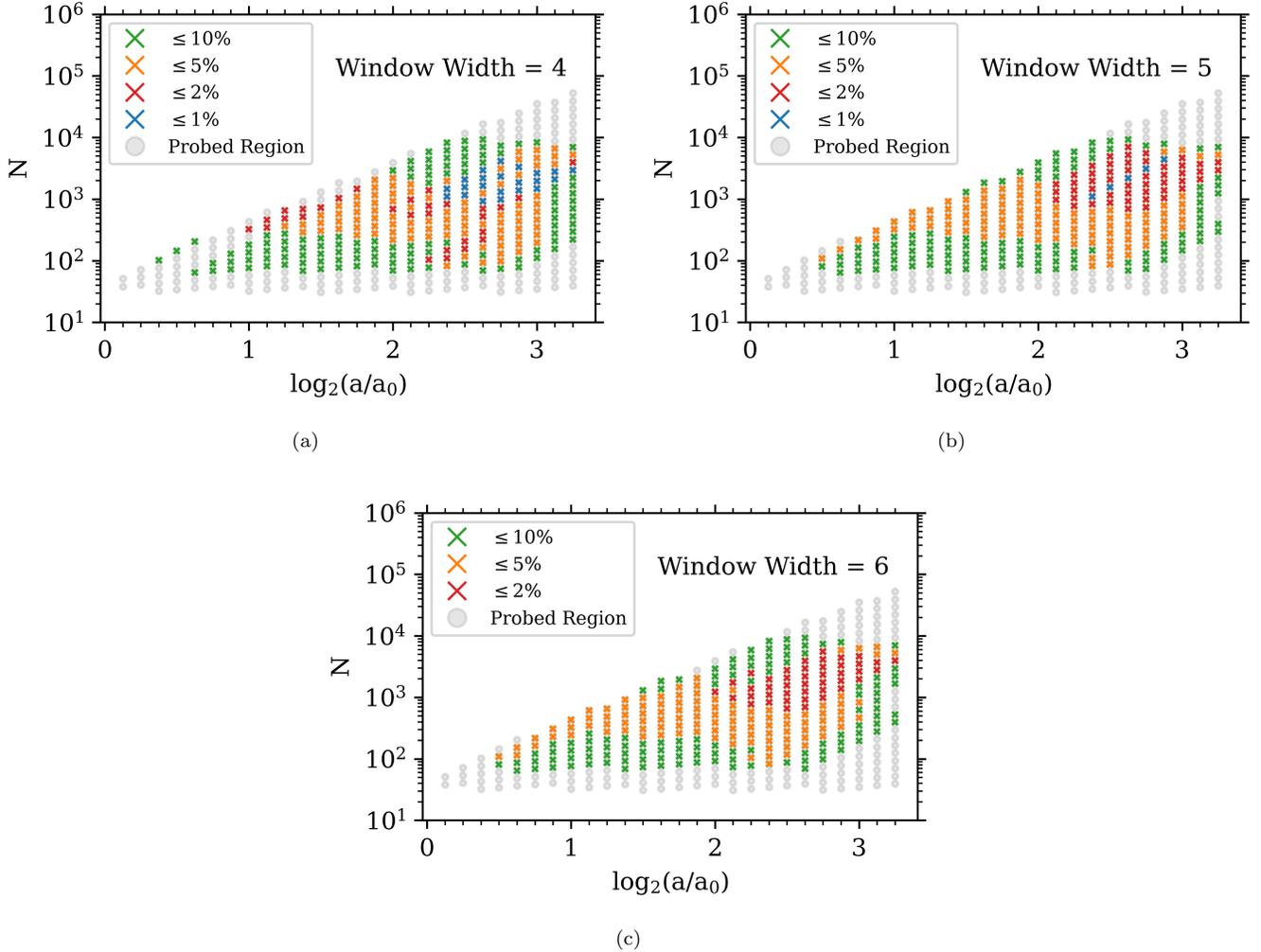

    \centering
    \null \vspace{-5pt}
    \gridline{\fig{Appendix_Plots_gtr20/CompaSO_n15_mean_past_joyce_mah_M200c_gtr20_FW_WW4_N_vs_a_convergence_MT2_UVM_with_text_min_1000.png}{0.49\textwidth}{(a)}
    \fig{Appendix_Plots_gtr20/CompaSO_n15_mean_past_joyce_mah_M200c_gtr20_FW_WW5_N_vs_a_convergence_MT2_UVM_with_text_min_1000.png}{0.49\textwidth}{(b)}}
    \gridline{\fig{Appendix_Plots_gtr20/CompaSO_n15_mean_past_joyce_mah_M200c_gtr20_FW_WW6_N_vs_a_convergence_MT2_UVM_with_text_min_1000.png}{0.49\textwidth}{(c)}}
    \caption{\raggedright Plots of the \textsc{CompaSO} convergence for different choices of window width in Equation \ref{eqn:delta1}. As discussed in Appendix \ref{varying_bin_width}, early time convergence is sensitive to changes in window width. The depth of convergence in the region of high convergence is also effected by reducing the window width size to four, with the four width results suggesting more 1\% convergence. The general location of this region remains the same regardless of the choice of window width, however. In all cases the absolute lower and upper bounds of convergence remain the same.}
    \label{fig:compaso_vary_window_N_v_a_convergence}
\end{figure*}

In Figures \ref{fig:rockstar_vary_window_N_v_a_convergence} and \ref{fig:compaso_vary_window_N_v_a_convergence} we compare the convergence results for different choices of $\Delta_1$ window width (see Section \ref{self_similarity}) for \textsc{Rockstar} and \textsc{CompaSO}. For both halo finders the general shape of the main region of convergence remains the same. Changes in convergence as a function of window width are most apparent at early times, where there are sometimes discontinuous regions of stability in the $\left<\Gamma\right>$ value (from Equation \ref{eqn:delta2}) as a function of time. This effect is most extreme when the size of the window is reduced, which creates islands of convergence for both halo finders. This sensitivity to our choice of metric suggests that this convergence is not robust. Moreover, as we argue in Section \ref{rockstar_results}, these early time regions of convergence should be treated with suspicion. Beyond these very early times, convergences regions are generally stable, with a reduction in window width size generating the most instability. This is unsurprising, given that with decreasing window width, our method becomes increasingly susceptible to noise.

\subsection{Tapered Tail Method}\label{tapered_tail}

\begin{figure*}
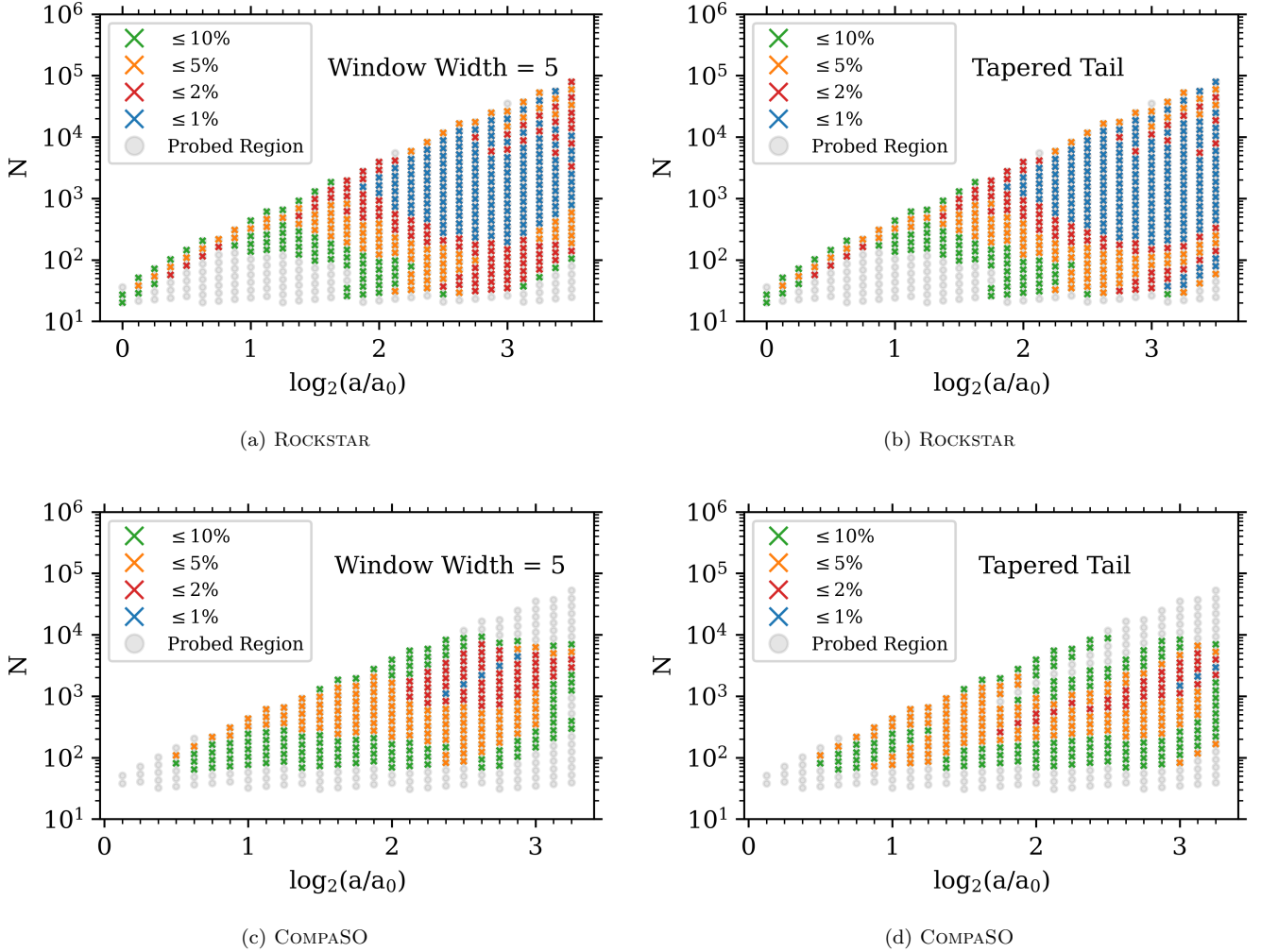

    \centering
    \null \vspace{-5pt}
    \gridline{\fig{Appendix_Plots_gtr20/Rockstar_n15_mean_past_joyce_mah_M200c_gtr20_FW_WW5_N_vs_a_convergence_MT2_UVM_with_text_min_1000.png}{0.49\textwidth}{(a) \textsc{Rockstar}}
    \fig{Appendix_Plots_gtr20/Rockstar_n15_mean_past_joyce_mah_M200c_gtr20_TT_WW5_N_vs_a_convergence_MT2_UVM_with_text_min_1000.png}{0.49\textwidth}{(b) \textsc{Rockstar}}}
    \gridline{\fig{Appendix_Plots_gtr20/CompaSO_n15_mean_past_joyce_mah_M200c_gtr20_FW_WW5_N_vs_a_convergence_MT2_UVM_with_text_min_1000.png}{0.49\textwidth}{(c) \textsc{CompaSO}}
    \fig{Appendix_Plots_gtr20/CompaSO_n15_mean_past_joyce_mah_M200c_gtr20_TT_WW5_N_vs_a_convergence_MT2_UVM_with_text_min_1000.png}{0.49\textwidth}{(d) \textsc{CompaSO}}}
    \caption{\raggedright Plots comparing our chosen convergence metric to the ''tapered tail" method discussed in Section \ref{tapered_tail}. The modified method allows us to probe lower particle counts at very late times, demonstrating the expected extension of convergence. It is more susceptible to \textsc{CompaSO}'s divergent behavior at high masses, however, and so we opt for our chosen method in the full in analysis.}
    \label{fig:tapered_tail_N_v_a_convergence}
\end{figure*}

One observed consequence of our choice of convergence metric was our inability to probe very late times as effectively. At very late times there are not enough snapshots in the lowest rescaled mass bins to probe convergence. By relaxing our window width requirement at very late times, we can accommodate smaller numbers of bins and potentially probe these very late very low mass halos. We did so by setting the window width of Equation \ref{eqn:delta1} equal to five snapshots, or the number of remain snapshots, whichever is smaller. Since we calculate $\Delta_1$ sequentially from early times to late times, the effect of this is to calculate $\Delta_1$ for the last four, three, and two sufficiently populated snapshots of a given rescaled mass bin. Because the last snapshot is by itself, it automatically passes the first stage of the convergence test. The remaining steps of the convergence test are the same, including the requirement that there be at least three consecutive of snapshots of convergence for a given rescaled mass bin. So the major change is that the $\left<\Gamma\right>$ value in Equation \ref{eqn:delta2} is now calculated using the mean mass accretion history of the last available snapshot, as well as potentially the second, third, and fourth to last, as well as any other five-wide snapshot regions that passed the first convergence check.

In Figure \ref{fig:tapered_tail_N_v_a_convergence} we plot the convergence regions for our current convergence metric (in (a) and (c)) and this modified ''tapered tail" method (in (b) and (d)). For \textsc{Rockstar}, this change of method has no major effect, but to continue the major region of convergence to lower particle counts at very late times. This very late time low particle improvement in convergence is repeated for the \textsc{CompaSO} results. \textsc{CompaSO}'s difficulty with very high mass halos causes problems for this method however. By relaxing the requirement window width at late times, we are effectively assuming that very late times always converge. This is true for all halos in \textsc{Rockstar}, but not true for large mass halos in \textsc{CompaSO} (see Figure \ref{fig:Rockstar_CompaSO_mah_v_a} for an example of \textsc{CompaSO}'s divergent behavior). When the window width is sufficiently reduced, the drop in mean mass accretion history for large mass halos at very late times is treated as a region of convergence. This dramatically skews the value of $\Delta_2$, obscuring any real convergence for smaller halos in that rescaled mass bin. In Figure \ref{fig:tapered_tail_N_v_a_convergence}, this is evident as a continuous break in the convergence region, extending from $\sim10^4$ particles per halo at $2^{2.5}\times a_0$ back down to $\sim10^3$ particles per halo at $2^{1.8}\times a_0$.
\subsection{Late Time Convergence}\label{low_z_convergence}
\begin{figure*}
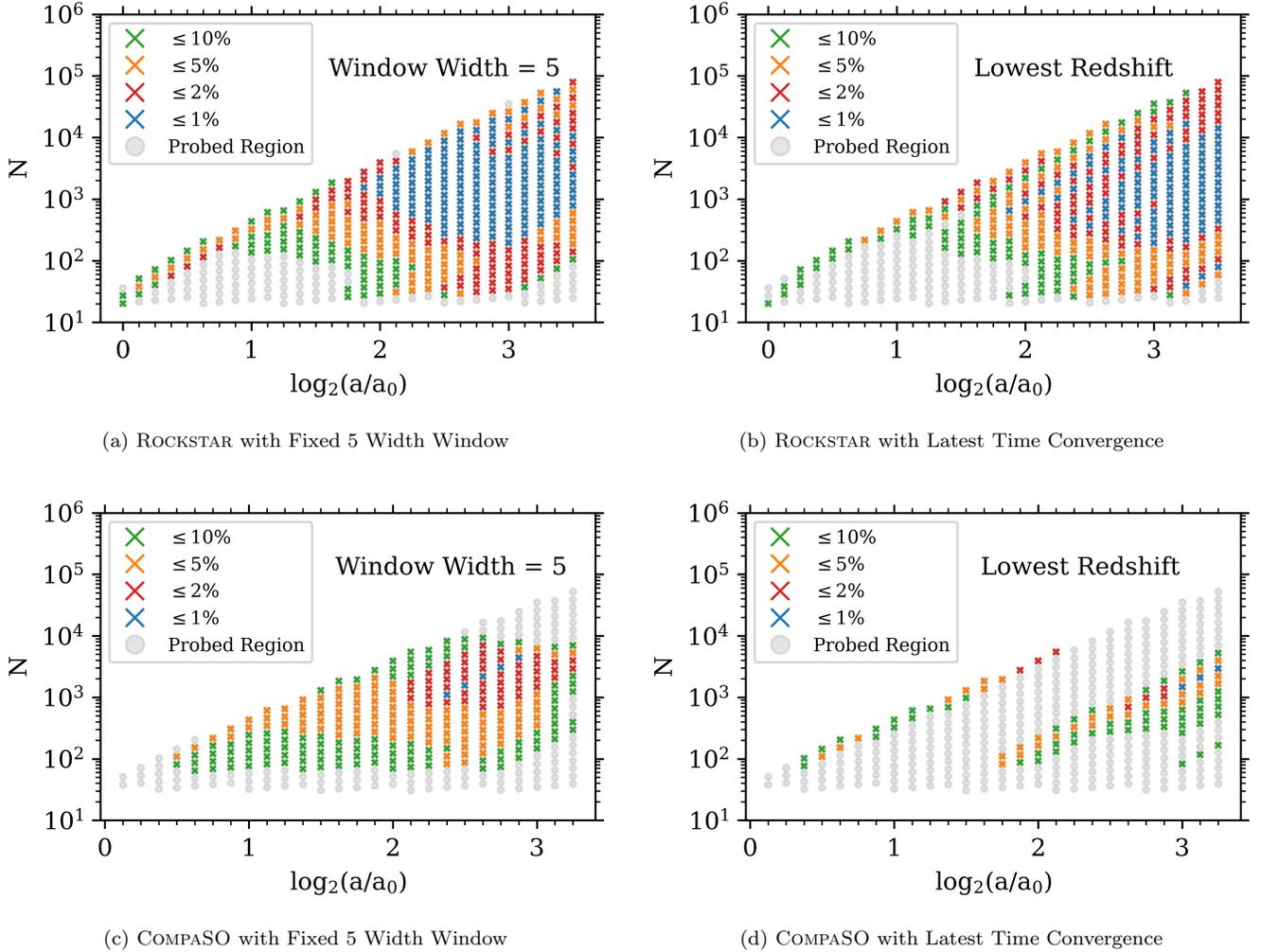

    \centering
    \null \vspace{-5pt}
    \gridline{\fig{Appendix_Plots_gtr20/Rockstar_n15_mean_past_joyce_mah_M200c_gtr20_FW_WW5_N_vs_a_convergence_MT2_UVM_with_text_min_1000.png}{0.49\textwidth}{(a) \textsc{Rockstar} with Fixed 5 Width Window}
    \fig{Appendix_Plots_gtr20/Rockstar_n15_mean_past_joyce_mah_M200c_gtr20_LZ_N_vs_a_convergence_with_text_min_1000.png}{0.49\textwidth}{(b) \textsc{Rockstar} with Latest Time Convergence}}
    \gridline{\fig{Appendix_Plots_gtr20/CompaSO_n15_mean_past_joyce_mah_M200c_gtr20_FW_WW5_N_vs_a_convergence_MT2_UVM_with_text_min_1000.png}{0.49\textwidth}{(c) \textsc{CompaSO} with Fixed 5 Width Window}
    \fig{Appendix_Plots_gtr20/CompaSO_n15_mean_past_joyce_mah_M200c_gtr20_LZ_N_vs_a_convergence_with_text_min_1000.png}{0.49\textwidth}{(d) \textsc{CompaSO} with Latest Time Convergence}}
    \caption{\raggedright Above is a comparison of convergence procedures. The left column of plots (a and c) show our default procedure described in Section \ref{self_similarity}, while the right column (b and d) shows the results of the procedure described in Section \ref{low_z_convergence}. The top row (a and b) displays results produced using the \textsc{Rockstar} halo finder, while the bottom row (c and d) displays the results produced using \textsc{CompaSO}. The difference between (a) and (b) is less immediately obvious than the difference between (c) and (d). Both (b) and (a) have large regions of high (1\%) convergence, but region is pushed further back in time for (b). This difference is likely caused by the increased sensitivity of this method to noise and slow changes in $\Gamma$. The procedure for the default convergence metric will naturally set the converged $\Gamma$ (i.e. $\left<\Gamma\right>$ in Equation \ref{eqn:delta2}) to the average of all snapshots that pass $\Delta_1 \leq X\%$. On the other hand, the latest time convergence metric effectively sets $\left<\Gamma\right>=\Gamma\left(a_{\rm{max}}\right)$. This means that in instances where there is a gradual transition to flatness in the $\Gamma$ vs $\log_2 (a/a_0)$ curve, or where the final snapshot has a slightly deviant $\Gamma$ value, the default method will be more likely to find weaker convergence across a large region, whereas the latest time method will by default favor a smaller later region (assuming that region includes $\Gamma\left(a_{\rm{max}}\right)$) to the potential exclusion of any earlier snapshots. The case in which the regions of self-similarity do not contain $\Gamma\left(a_{\rm{max}}\right)$ is illustrated in the \textsc{CompaSO} results shown in (d). There, because of divergent behavior at late times, $\Gamma\left(a_{\rm{max}}\right)$ is rarely in a region of flatness. Therefore, using $\Gamma\left(a_{\rm{max}}\right)$ as a baseline, one rarely finds any convergence at all. It is exactly this flaw that dissuaded us from using this method in the main analysis, but we include it for completeness.}
    \label{fig:low_z_convergence}

\end{figure*}

The last extension to our convergence metric that we considered was to do away with $\Delta_1$ entirely. Instead, we treated the mean mass accretion history of the latest sufficiently populated snapshot as the converged mean mass accretion history ($\left<\Gamma\right>$ in Equation \ref{eqn:delta2}). We then followed the same procedure (i.e., still requiring $\Delta_2 < X\%$ and still requiring three consective snapshots for a region of convergence). The results of this modified procedure are shown in Figure \ref{fig:low_z_convergence}. As expected, this procedure works well for \textsc{Rockstar}, where the mean mass accretion history of a rescaled mass bin converges to a single value as the scale factor increases. Equally as unsurprising, at least in light of the results in Section \ref{tapered_tail}, this result does not work at all for \textsc{CompaSO}. \textsc{CompaSO}'s difficulty with halos with $\gtrsim 10^4$ particles is amplified by this procedure. Because $M_{\mathrm{NL}}$ increases with time, the latest available snapshots of a given rescaled mass bin are also those containing the most massive halos. When these halos are sufficiently large, their mean mass accretion history as calculated by \textsc{CompaSO} diverges. Since this modified method uses these values as its benchmarks for convergence, this behavior totally obscures any convergence in rescaled mass bins where the halos get too large.

\end{document}